
\input phyzzx
\overfullrule=0pt
\def\n{\noindent}
\def\p{\partial}

\def\vp{\hat \varphi }
\def\hpi{\hat\pi }
\def\an{\alpha_n}
\def\ga{\alpha}

\null
\rightline {UTTG-11-93}
\rightline {March 1993}

\title{Discrete Strings and Deterministic Cellular Strings
\foot{Work supported in part by NSF grant PHY 9009850 and
R.~A.~Welch Foundation.}}

\author{Jorge G. Russo }
\address {Theory Group, Department of Physics, University of
Texas\break
Austin, TX 78712}

\abstract
 A discrete string theory --a theory of embeddings from ${\bf Z}\times
{\bf Z}_C\to {\bf R}^D$, where $C$ is the number of components of the
string-- is explored.
The closure of the algebra of constraints (`${\bf Z}_C$-Virasoro algebra')
is exhibited.
The ${\bf Z}_C$-Virasoro `algebra' is shown to be anomaly free
in arbitrary number of target space dimensions.
We prove the existence of a (manifestly unitary) light-cone gauge
with anomaly free Lorentz algebra in any dimensions.
The analog of
vertex operators are introduced and the physical spectrum is analysed.
There are an infinite number
of higher-level states repeating a certain mass pattern and leading to an
infinite degeneracy. The connection with the continuum string theory
(in $D=26$) is investigated.
Independently, following a method recently introduced by 't Hooft based
on Hilbert-space extension of deterministic systems, a particular
one-dimensional cellular automaton submitted to a deterministic evolution
is shown to reduce to a massless scalar field theory at long distance scales.
This automaton is utilized to define a (`cellular') string theory
with world-sheet variables evolving under deterministic rules, which
in the framework of `first-quantization' corresponds to
the ${\bf Z}_C$ string theory mentioned above.
We show that in this theory  also the target space motion of
free strings is governed by deterministic laws. Finally,  we discuss
a model for (off-shell) interacting strings
where space-time determinism is fully restored.

\endpage

\Ref\fey {R. P. Feynman, Int. J. Theor. Phys. 21, (1982) 467.}

\Ref\hoof{G. 't Hooft, {\it On the quantization of space and time},
in {\it Proceedings of the Moscow Seminar
on Quantum Gravity} (Academy of Sciences of the USSR, 1987).}

\Ref\hooft {G. 't Hooft, J. Stat. Phys. 53, (1988) 323;
Nucl. Phys. B342 (1990) 471.}

\Ref\hooftt {G. 't Hooft, K. Isler and S. Kalitzin, Utrecht preprint,
THU-92/07.}

\Ref\ks {I. Klebanov and L. Susskind, Nucl. Phys. B309
(1988) 175.}

\Ref\aw {J.J. Atick and E. Witten, Nucl. Phys. B310
(1988) 291.}

\Ref\veselov {S.P. Novikov, Funkts. Anal. Prilozhen. 8 (1974) 54;
A. G. Reyman and M. A. Semenov, Invent. Math. 54 (1979) 81;
W.W. Symes, Physica 4D, (1982) 275;
A.P. Veselov, Funkts. Anal. Prilozhen., 22 (1988) 1.}

\Ref\gsw {M. Green, J. Schwarz and E. Witten: {\it Superstring Theory},
Cambridge Univ. Press, Cambridge (1987).}

\chapter{Introduction}

The reduction to discrete variables or to {\it simpler} systems
has always represented an objective in physics.
Numerous physicists have attempted to write
the final chapter of discretization, namely the
quantization of space and time. Some believe that it is at this level
that the determinism in physics should be restored.
In particular, in ref. [\fey ] Feynman
expressed his insatisfaction with our present understanding of
quantum mechanics and speculated that
at more fundamental level Nature might resemble
a cellular automaton (i.e., an array of elements or {\it cells}
with associated dynamical variables which, at each time step,
 evolve according to a given law).

An important step in this direction was recently performed
by  't Hooft.
{}From his studies on
the quantum physics underlying black hole
geometries,  't Hooft argues that the number of mutually orthogonal
states in Hilbert space inside a closed surface
is finite and given by the total area of the surface [\hoof ].
This suggests that at the Planck scale space-time
may be discrete.
A subclass of discrete theories are, precisely, the
deterministic discrete theories, in which there
is a basis ({\it primitive basis}, in 't Hooft terms),
where the evolution matrices $U(t_i,t_j)$ are pure permutation matrices.
At successive times $t_1, t_2, t_3,...$ the basis elements are just
permuted, $|e_1\rangle\to |e_2\rangle \to |e_3\rangle\to ...$
. For an arbitrary vector $|\psi \rangle $ (a linear superposition
of $|e_i \rangle $) one defines the probability of a given state
$|e \rangle $ as [\hooft ]:
$$
P(e)=|\langle e|\psi \rangle |^2\ .
\eqn\probab
$$
It turns out that there is a natural way to reproduce the evolution of
$|e\rangle $ by writing a Schr\" odinger equation for $|\psi \rangle $:
$$
{d\over dt}|\psi\rangle =-i H|\psi \rangle\ .
\eqn\shod
$$
At {\it integer} time $t$ any basis element $|e\rangle $ evolves into
$|e_t \rangle $ in accordance to the law of the cellular automaton, but
eq. \shod\ also prescribes how the phase factors evolve.

 't Hooft suggests that  quantum mechanics  may be viewed
{\it not as a theory about reality, but as a prescription
for making the best possible predictions about the future if
we have certain information about the past.}
The Einstein-Podolsky-Rosen paradox, Bell's theorems, {\it Gedanken}
experiments, etc. show that one cannot attach labels to electrons
to determine their evolution, but they do not prove that
hidden variables associated with the vacuum cannot restore predicability
in a formal way. As a matter of fact, 't Hooft provides explicit
cellular automaton examples
where determinism is restored in virtue of hidden variables
associated with the vacuum [\hooft ].
 These automata mostly evolve chaotically; their
long-distance behaviour can only be treated by using statistics.
If in some case an effective quantum field theory emerges
with a large fundamental distance scale, at large scales
only renormalizable (and super renormalizable) couplings survive,
thus one of the renormalizable theories should be reproduced.

An example
is  the Fermi one-dimensional shift automaton [\hooft, \hooftt]
(in sect. 2 we will introduce the bosonic analog).
One divides the circle in $C$ cells, $C=2N+1$, where $N$ is a natural
number.
There is a
a variable $\sigma_x$ which in each cell can take the values $0$
or $1$. The evolution equation is taken to be
$$
\sigma _{x,t+1}=\sigma _{x-1,t}\ \ ,\ \ \ \ \sigma_{C,t}
=\sigma_{0,t}\ \ ,
\eqn\spines
$$
i.e. the spins shift to the right at constant speed. The Hilbert space
is generated by the basis elements $|\sigma_1,...,\sigma_C\rangle $.
Having zeros and ones, one may introduce  anticommuting
creation and annihilation operators $\hat \sigma ^\pm _x$ at each
site $x$, with
$$
\hat \sigma^-_ x |\sigma_x= 0\rangle =0\ \ ,
\ \ \ \ \hat \sigma^-_ x |\sigma_x= 1\rangle =
|\sigma_x= 0\rangle \ \ ,
\eqn\aniq
$$
whereas the operation of the $\sigma^-_x$ operators does not depend
on the contents of the other cells.
Then one applies a Jordan-Wigner transformation
$$
\psi_x =(-1)^{\sum_{y<x} \sigma_y} \ \hat\sigma^-_x\ \ ,
\eqn\jw
$$
so that
$$
\{ \psi_x, \psi_y\} =0\ ,\ \ \ \{ \psi_x, \psi_y^{\dag } \} =\delta_{xy}
\ \ ,
\eqn\commi
$$
and
$$\psi_{x,t}=\psi_{x-t,0}\ .$$
By Fourier transforming
$$
\psi _{x,t} ={1\over \sqrt{C}}\sum _{k=-N}^N \tilde \psi _{k,t}\
e^{-2\pi ikx/C}\ \ ,
\eqn\foury
$$
one learns that the evolution equation \spines\ is generated by
a Hamiltonian operator,
$$
H={2\pi\over C}\sum_{k=-N}^N k\tilde \psi^{\dag }_{-k}\ \tilde \psi_k
\ \ ,
\eqn\jamil
$$
which in the continuum limit approaches
to
$$
\int dx \ \psi^{\dag }(x)i {\p\over \p x} \psi (x)
$$
i.e. the Hamiltonian of a chiral, right-handed fermion.
The ground state $|0\rangle $ is obtained by filling the Dirac sea with
negative-$k$ modes. This state is a linear combination
of all states $|\sigma _1,...,\sigma_C \rangle  $ of the
primitive basis.
Physical particles are the excitations
above this lowest energy state.

A common problem that arises in discrete-time scenarios and it
is of course present here is that {\it energy} can only
be defined modulo $2\pi $. We lack a clear physical
understanding on how this may be resolved.
Another delicate point is the continuum limit.
If we send $C\to\infty $ before introducing a regulator
for continuum expressions, for example, in building
$:\psi^{\dag }\psi :$,
we miss  the correct Schwinger anomaly in the
corresponding current algebra.
The reason is a cancellation due to contributions of
states of momentum $k\sim \pm N$. One would like
to keep only low-energy excitations of the field $\psi $
in the process of taking the continuum limit. This is
automatically accomplished if a proper regularization is introduced
prior to letting $N\to\infty $.

In the last decade string theories  have fed an
unusual hope for a consistent
unification of all forces of Nature including gravity.
Despite some remarkable progresses, unfortunately
string theory continues being
no more than a set of Feynman rules.
Confined to a first-quantized formalism, very little
can be learned about its origins,
and few and weak low-energy predictions can be unambiguously
stated.
The theory is incomplete and one is urged to the
search for fundamental principles.

It is tempting to compare with the large-$N$ QCD analogy.
Below the deconfining transition effective strings and
continuum Riemann surfaces
constitute a correct description of large-$N$ QCD, but above
the deconfining transition the Riemann surface picture is totally
inadequate. Riemann surfaces are inundated by a
sea of holes and they
must be replaced by Feynman diagrams.
Today we know that quantum cromodynamics is
the correct fundamental description.

In the case of string and superstring theories,
there have been some dim indications that
a more fundamental theory should have some discrete
structure.

In particular, in ref. [\ks ] it was shown that one of the phases of a
light-cone lattice gauge theory with an infinite number of colors
 describes free fundamental strings, even if the lattice spacing
is not taken to zero.

The fact that a continuum Riemann surface description  must break down
is evidenced by the presence of the Hagedorn transition.
Above the Hagedorn temperature the free energy appears to have
a genus zero contribution [\aw ], which implies that the world
sheet is no longer simply connected. Indeed, at $T=T_H$ a solitonic
mode becomes tachyonic and develops a vacuum expectation
value, whereby creating a conglomeration of holes in the Riemann surface.
The conclusions of ref. [\aw ] are that continuum world sheets
should be replaced by some less continuum structure, and
that the density of gauge invariant degrees of freedom
of string theories is much less than any ordinary relativistic
field theory.
A problem which is quite
analogous to the $2\pi $ ambiguity in the energy
of the cellular automata is the duality relation $T\to 1/T$ in
the heterotic string. The consequences of this symmetry, if preserved,
would be catastrophic, since this symmetry implies that
the thermodynamical partition function $Z$ at infinite
temperature is equal to $Z(T=0)$, which  would mean
 that there is no fundamental gauge-invariant degrees of freedom
 in the theory. The expectation is that  this duality
symmetry is spontaneously broken.

This work is organized as follows. In sect. 2 we
introduce a one-dimensional cellular automaton
subject to a simple deterministic law and demonstrate that, in the
large-distance regime, this automaton
can be described in terms of a quantum field theory of a massless
scalar field moving in $1+1$ dimensions.
The ground state is expressed in terms of the {\it primitive} basis
in a non-trivial way.
In sect. 3 we introduce the open and closed
${\bf Z}_C$ discrete string theories.
The algebra of the constraints is studied and interpreted.
 Sect. 4 contains the calculation of the ${\bf Z}_C$-Virasoro
anomaly. The light-cone gauge is investigated in sect. 5.
In particular, we show that Lorentz covariance is maintained
in any number of space-time dimensions.
In sect. 6 the spectrum is analysed both in the light-cone
gauge formalism and in the covariant approach.
Section 7 contains a discussion on vertex operators, and
sect. 8 deals with the connection with the continuum
theory and the scattering amplitudes.
In sect. 9
we consider the string theory
that results upon placing the cellular automaton of sect. 2
to dictate the world-sheet dynamics. The deterministic evolution
of free cell strings in target space is disclosed.
We argue that deterministic laws governing the
evolution of an arbitrary number of {\it interacting} cell strings,
describing in particular splitting and joining, are possible.
To elucidate this, we introduce a model for a fully deterministic cellular
string theory. Given an arbitrary initial configuration of closed
an open strings, the whole evolution, including possible splitting
or joining of the strings and distribution of momenta of emerging strings,
is determined by very simple rules.
However, in cases  where the number of cells in play is large,
the evolution of the system becomes so complex that it
can be more suitably studied by  numerical or statistical methods.
In sect. 10 discuss general aspects of the theory
presented in the main text. In addition, some speculations are made
on how this theory could
find application in realistic models.

\bigskip

\chapter {The Bosonic Cellular Automaton}

Let us consider a partition of the circle in $C$ cells.
In each cell $x$ there are two variables
$V^L_x, V^R_x, \ x=1,...,C$ which take values in ${\bf R}$.
The evolution equation is defined by

$$
V^R_{x,t+1}=V^R_{x-1,t}\ \ ,\ \ \ \ V^L_{x,t+1}=V^L_{x+1,t}\ \ ,
\eqn\auto
$$
$$
V^R_{0,t}=V^R_{C,t}+{q\over 2}\ ,\ \ \ V^L_{C+1,t}=V^L_{1,t}+{q\over 2}\ ,
\ \ \ \ \ \ \ q={\rm constant}\ .
\eqn\bordes
$$
Using that
$$
V^L_{0,t}=V^L_{1,t-1}=V^L_{C+1,t-1}-{q\over 2}=V^L_{C,t}-{q\over 2}
$$
one sees that the sum $V_x^R+V_x^L$ is single-valued:
$$
(V^R+V^L)_{C,t}=(V^R+V^L)_{0,t}\ \ .
\eqn\single
$$
The `zero mode' part
$$
v_t^0={1\over C}\sum_{x=1}^C(V^L_{x,t}+V^R_{x,t})
\eqn\zerom
$$
will be subject to the law
$$
v^0_{t+1}=v^0_t+{1\over C}q \ .\ \ \ \ \
\eqn\lawzero
$$
Separating the zero mode part we can describe the automaton in terms
of single-valued variables $v^R_{x,t}, v^L_{x,t}$, i.e. by writing
$$
V^L_{x,t} +V^R_{x,t}=v^0_t+v^L_{x,t}+v^R_{x,t}\ ,\ \ \
\eqn\cambios
$$
$$
v_{x,t}^{R,L}=V^{R,L}_{x,t}-(t\mp x)
{q\over 2C}+ {\rm const.}\ \ ,
\eqn\nuevasv
$$
where $v^R_{x,t}, v^L_{x,t}$ evolve  as follows
$$
v^R_{x,t+1}=v^R_{x-1,t}\ \ ,\ \ \ \ v^L_{x,t+1}=v^L_{x+1,t}\ \ ,
\ \ \ \ \ \ v^{R,L}_{x+C,t}=v^{R,L}_{x,t}\ \ .
\eqn\autom
$$

The Hilbert space is spanned by the basis
$$
\{ |v^0\rangle \otimes|v_1^L,...,v^L_C\rangle
\otimes |v_1^R,...,v^R_C\rangle \} \ .
\eqn\base
$$

Let us introduce operators $\vp _x, \hpi_x $
at each site $x$ with $[\vp_x,\hpi_y]=i\delta_{xy}\ $,
 and the decomposition
$$
\vp _x=\vp^L _x+\vp^R _x +\vp^0 \ ,\ \ \ \vp^0={1\over C}
\sum_{x=1}^C\vp_x\ ,
\eqn\deco
$$
$$
\hpi _x=\hpi^L _x+\hpi^R_x+{1\over C}\hpi^0\ ,\ \ \ \hpi^0=
\sum_{x=1}^C\hpi_x\ ,
\eqn\decopi
$$
satisfying
$$
[\vp_x^R,\hpi_y^R]=[\vp_x^L,\hpi_y^L]={C-1\over 2C}i \delta_{xy}\ ,\ \
[\vp^0,\hpi^0]=i \ \ ,
\eqn\commu
$$
and other commutators equal to zero.

The eigenstates of $\vp^L_x, \vp^R_x, \vp^0$ are given by
$$
|v \rangle \equiv |v^0\rangle\otimes
|v_1^L,...,v^L_C\rangle\otimes |v^R_1,...,v^R_C\rangle =
e^{-iv^0 \hpi^0 }\prod_ {x=1}^C e^{-i{2C\over
C-1}(v_x^L\hpi^L_x+v_x^R\hpi^R_x)}
|v=0\rangle \ ,
\eqn\eige
$$
where
$\vp_x |v=0\rangle =0$. It is easy to verify that
$$
\vp^L_x|v\rangle =v^L_x|v\rangle\ ,\ \
\vp^R_x|v\rangle =v^R_x|v\rangle\ ,\ \ \vp ^0|v\rangle =v^0 |v\rangle \ .
\eqn\valu
$$

Let us now expand $\vp^R_x, \vp^L_x$ in Fourier components
$$
\vp^R_x={i\over 2}l\sum_{n=1}^N{1\over \sqrt{n}}\big(
a_n\omega^{nx}-a_n^{\dag } \omega^{-nx}\big)\ \ ,\ \ \
\ \vp^L_x={i\over 2}l\sum_{n=1}^N{1\over \sqrt{n}}\big(
\bar a_n\omega^{-nx}-\bar a_n^{\dag } \omega^{nx}\big)\ ,
\eqn\righ
$$
$$
\omega\equiv e^{i{2\pi\over C}}\ ,
$$
where $N=(C-1)/2$ if $ C$ is odd, $N={C\over 2}-1$ if $C$ is even,
and $l$ is a constant with dimension of length introduced
to match standard string theory conventions in the continuum limit
(see below).
For more simplicity in the presentation of the formulas,
in part of this work we will only refer explicitly to one parity of $C$,
namely, $C$ odd $\equiv 2N+1$, the treatment of the other case being
exactly the same.

The momentum operator can be defined in terms of the
modes $a_n,a_n^{\dag },\bar a_n, \bar a_n^{\dag }$ as follows
$$
\hpi^R_x={1\over Cl}\sum_{n=1}^N \sqrt{n}\big(
a_n\omega^{nx}+ a_n^{\dag } \omega^{-nx}\big) \ ,
\eqn\momri
$$
$$
\hpi^L_x={1\over Cl}\sum_{n=1}^N \sqrt{n} \big(
\bar a_n\omega^{-nx}+ \bar a_n^{\dag } \omega^{nx}\big) \ .
\eqn\momle
$$
{}From the canonical commutation relations for $\vp$ and $\hpi $
it follows
$$
[a_n,a_m^{\dag }]=[\bar a_n,\bar a_m^{\dag }]=\delta_{nm}\ \ ,
\eqn\canon
$$
and the remaining commutators vanish.

In the Heisenberg picture we have
$$
\vp^L_{x,t}=\vp^L_{x+t,0}\ ,\ \ \vp^R_{x,t}=\vp^R_{x-t,0}\ \ ,
$$
$$\vp^0_t=\hat \phi+ {\pi l^2\over C}p t\ ,\ \ \
\ \hat\phi\equiv\vp^0_{t=0}\ ,\ \ \
p\equiv {1\over \pi l^2} q\ .
\eqn\heisen
$$

It will turn convenient introducing the operators
$$
\eqalign{\an &=\sqrt{n}a_n\ \ , \ \ \an^{\dag }=\ga_{-n}=\sqrt{n}a^{\dag }
_n \ ,
\cr
\bar\an &=\sqrt{n}\bar a_n\ \ , \ \ \bar\an^{\dag }=\bar\ga_{-n}=\sqrt{n}
\bar a^{\dag }_n\ ,\ \ \  n=1,...,N\ .\cr }
\eqn\alfas
$$
{}From eqs. \righ,  \heisen\ and \alfas\
we see that the time evolution for $\an\ ,\ \bar\an $ is given by
$$
\ga_{n,t}=U^{\dag } (t)\ga_{n,0} U(t)=\omega^{-nt}\ga _{n,0}\ ,\ \ \
\bar \ga_{n,t}=U^{\dag } (t)\bar \ga_{n,0} U(t)=\omega^{-nt}\bar \ga _{n,0}\ .
\eqn\alfatime
$$
Therefore we take as Hamiltonian
$$
H={\pi l^2\over C} p\hpi^0 +{2\pi\over C}\sum _{n=1}^N
(\ga_{-n}\an +\bar\ga_{-n}\bar\an ) +{\rm const.}
\eqn\hamil
$$

The continuum limit is taken by rescaling $t,x\to t/ d, x/ d  $,
and letting $N\to \infty $, $d\to 0$ keeping the lenght of the circle
$Cd$ finite, $Cd=\pi $.
 In this limit $\vp $ becomes
$$
\vp =\hat \phi +l^2pt+{i\over 2}l\sum_
{{n=-\infty}\atop n\neq 0}^\infty {1\over n}
(\an e^{-2inx^-} +\bar\an e^{-2inx^+})\ ,\ \ \ x^\pm=t\pm x\ .
\eqn\contin
$$
This resembles the scalar field in one time plus one compact-space
dimensions of
string theory, with $p$ representing the center of mass target momentum
of the string and $l^2\equiv 2\ga '$. However, there is an important
difference:
because we have requested the left and right moving components to transform
by a given number $p$ (see eq. \bordes ), in eq. \contin ,
instead of the  operator $\hpi^0$, the constant $p$ appears.
The difference can also be appraised in the zero-mode structure of the
automaton Hamiltonian, where $p\hpi^0 $ replaces the usual
$(\hpi^0)^2$ of the string-theory Hamiltonian.

The vacuum state is defined by
$$ \hpi^0 |0\rangle = 0\ ,\ \ \
\an |0\rangle =\bar \an |0\rangle =0\ , \ \ \ n=1,...,N \ ,
\eqn\vacuum
$$
and
$$
\langle 0| \hpi^0  = 0\ ,\ \ \
\langle 0|\an =\langle 0|\bar \an =0\ , \ \ \ n=-1,...,-N \ .
\eqn\vacuumd
$$
This state is obviously stationary in time. The
physical two-dimensional particles are the excitations
above the ground state $|0\rangle $.
The
state $|v^L=0, v^R=0 \rangle $
is obtained from $|0\rangle$ by the following formula
(we omit the usual relation between $|\pi^0=0\rangle$
and $|v^0=0\rangle $)

$$
|v=0\rangle =\prod_{n=1}^N \exp[{1\over 2}
a_n^{\dag } a_n^{\dag } \omega^{-2nx}]\exp[{1\over 2}
\bar a_n^{\dag } \bar a_n^{\dag } \omega^{2nx}] |0\rangle \ \ .
\eqn\vcero
$$
To see this it is sufficient to consider a single oscillator.
We have to show that
$$
(a c-a^{\dag }{\bar c})e^{{1\over 2}a^{\dag 2}{\bar c}^2}|0\rangle =0\ ,
\ \ \ \ c{\bar c}=1\ \ .
\eqn\unosci
$$
By expanding the exponential function and using the algebra \canon\
and the definition of the vacuum $|0\rangle $, eq. \vacuum ,
we obtain the result stated above:
$$
\eqalign
{(a c-a^{\dag }{\bar c})e^{{1\over 2}a^{\dag 2}{\bar c}^2}|0\rangle &=
\sum_{n=1}^\infty {{\bar c}^{2n-1}\over (n-1)!2^{n-1}}a^{\dag 2n-1}|0\rangle
-\sum_{n=0}^\infty {{\bar c}^{2n+1}\over n!2^n} a^{\dag 2n+1}|0\rangle \cr
&= 0 \ \ . \cr }
\eqn\prova
$$
Using eq. \vcero \ one can compute the 't Hooft `quantum probabilities'
$P(v)=|\langle v|0\rangle|^2$ mentioned in sect. 1.

{}From the mode operator algebra and the definition of the vacuum state
it is easy to compute the correlation functions. We obtain
$$
\langle \vp_{x,t}^R\vp_{x',t'}^R \rangle=
{l^2\over 4}\sum_{n=1}^N
{1\over n} \omega^{-n(x^- -x'{ }^-)} \ \ ,
\eqn\correr
$$
$$
\langle \vp_{x,t}^L\vp_{x',t'}^L \rangle=
{l^2\over 4}\sum_{n=1}^N
{1\over n} \omega^{-n(x^+ -x'{ }^+)}\ \ .
\eqn\correl
$$
Note that these correlators are finite at coinciding points.
In the continuum limit they exactly reproduce
the well-known logarithmic expression.

The `open' bosonic cellular automaton is obtained by imposing
reflecting boundary
conditions on the end cells. Let the cell string be given by
$x=0,1,2,...,N+1$ with variables $v^{R}_x, v^{L}_x$ associated with
each cell and the zero mode variable $v^{0}$ .
The automaton rules are the following:
$$
v^{R}_{x, t+1}=v^{R}_{x-1, t}\ ,\ \ \ \  x=1,...,N+1\ \ ,
\eqn\derecha
$$
$$
v^{L}_{x, t+1}=v^{L}_{x+1, t}\ ,\ \ \ \ x=0,1,...,N\ \ ,
\eqn\zurda
$$
$$
v^{0}_{t+1}=v^{0}_{t}+{1\over N+1}q\ ,\ \ \ \ \ q={\rm constant}\ ,
\eqn\autopen
$$
$$
v^{R}_{0, t}=v^{L}_{0, t}\ ,\ \ \ \ \
v^{L}_{N+1, t}=v^{R}_{N+1, t}\ ,
\eqn\bautopen
$$
{}From eqs. \derecha , \zurda , \bautopen\ it follows that
$v^{R}_{0, t+1}=v^{L}_{1, t}$ and
$v^{L}_{N+1, t+1}=v^{R}_{N, t}$.

To obtain the effective long-distance field theory we proceed
in the same way as we did in the closed case,
that is, we
introduce operators $\vp _x, \hpi_x $
at each site $x$ with $[\vp_x ,\hpi_{y}]=i\delta_{xy} $,
etc. The Fourier expansion which satisfies the boundary conditions
\bautopen\ is given by
$$
\vp _{x,t}=\hat \phi +\beta p t
+i{l\over 2} \sum_{ n=-N \atop n\neq 0 }^N
{1\over n}\ga_n (\xi ^{-nx^-}
+\xi ^{-nx^+})\ ,
\eqn\vpopen
$$
where
$$
\xi=e^{i\pi\over N+1}\ ,\ \ \ \ \ \ \ \beta={\pi l^2\over N+1}
\ , \ \ \ p={1\over \pi l^2}q\   .
$$

As a result, $\vp _x$ is periodic with period $2(N+1)$,
$\vp _{x+2N+2}=\vp _x$.
The commutation relations for the $\an $ are the same as in the
closed case (see eqs. \alfas , \canon )
$$
[\an , \ga_m  ]=n \delta_{n+m} \ .\
\eqn\alfasene
$$

In a similar way as in the previous
case,
the continuum limit is taken by rescaling $t,x\to t/ d, x/ d  $,
and letting $N\to \infty $, $d\to 0$ keeping the lenght
$(N+1)d$ finite, $(N+1)d=\pi $. The resulting field theory is
that of a conformal scalar field satisfying Neumann
boundary conditions with a zero-mode constrained by eq. \autopen .
\bigskip

\chapter { Discrete String Theory }

It is straightforward to extend the previous model to $D$ scalar
fields $\vp^\mu $ by allowing for $2D$ variables $V_x^{L\mu }  , \
V_x^{R\mu }\ ,\ \   \mu=0,1,...,D-1$, at each site
$x$, $x=1,...,C$,  which evolve according
to the deterministic rules eqs.  \auto , \bordes .
As will be discussed in sect. 9, in this {\it cellular string theory}
free cellular strings evolve under deterministic rules in target space.
The basic reason is that the target-space momentum $p^\mu$ is a c-number
and hence it commutes with the center of mass coordinate.
In the 't Hooft scenario, the reconciliation with
target-space quantum mechanics
should be achieved by applying the same method that we applied in
the world sheet to {\it target} space-time. That is, one defines
the Hilbert space of second quantization and looks for a primitive basis
where the evolution matrices are just permutation matrices.
The objects that we would call `particles' or
physical states would be an intrincated combination
of the primitive basis elements. In principle, there is no
reason why Bell inequalities for the primitive basis elements
should imply Bell inequalities for the physical states.
An obstacle to accomplish this strategy is the absence
of a second-quantized string theory. Furthermore, there is a previous step
which has not been performed yet, namely
finding a cellular
automaton which at large scales describes
a massless scalar field theory in $D>2$ dimensions
(the $D=2$ case is the discussion of sect. 2).
We will return to the deterministic cellular string theory in sect. 9.
Meanwhile, we come back to the formalism of first-quantization
and promote the constant $p^\mu$ to an operator $\hat p^\mu$. We have
(cf. eqs. \righ , \heisen , \alfas )
$$
\vp ^{\mu}_{x,t}=\hat \phi^\mu +{\pi l^2\over C}\hat p^\mu t
+i{l\over 2}\sum_{ n=-N \atop n\neq 0 }^N{1\over n}
(\ga_n^\mu\omega ^{-nx^-}
+\bar\ga_n^\mu\omega ^{-nx^+})\ ,\ \ \ \ \omega \equiv
e^{i{2\pi\over C}} \ \ .
\eqn\vpstring
$$
Aside the zero-mode part, the left and right moving components can still
be interpreted in terms of the cellular automaton of sect. 2.

An alternative interpretation
is saying that we do not want deterministic evolution and investigating
discrete string theories
for their own sake, i.e. replace
the world-sheet by the lattice ${\bf Z}\times {\bf Z}_C$, introduce
the operator \vpstring\ and its canonical conjugate $\hpi _{x,t}^\mu$
acting on the Fock space, etc. Sects. 3 to 8 apply to either
interpretations.

The operator $\vp^\mu $ with $\mu=0$ will be identified with
time coordinate in target space so we shall reverse the signs in
the appropriate commutators. The nonvanishing
commutation relations for the oscillator mode operators are
(see eqs. \canon , \alfas )
$$
[\an ^\mu, \ga_m ^\nu ]=n \delta_{n+m}\eta^{\mu\nu }\ ,\ \ \
[\bar \an ^\mu, \bar \ga_m ^\nu ]=n \delta_{n+m}\eta^{\mu\nu }\ ,\ \ \ \
[\hat \phi ^\mu , \hat p^\nu ]=i\eta^{\mu\nu } \ ,
\eqn\modesc
$$
where $\eta_{\mu\nu } $ is the Minkowski metric with signature
$\{ -+...+\} $. Similarly, the vacuum state is defined by
$$\hat p^\mu |0 \rangle=0 \ ,\ \ \
\an ^\mu |0\rangle =\bar \an ^\mu |0\rangle =0\ , \ \ \ n=1,...,N \ ,
\eqn\gvacuum
$$
and
$$
\langle 0|\hat p^\mu =0 \ ,\ \ \
\langle 0|\an ^\mu =\langle 0|\bar \an ^\mu=0\ , \ \ \ n=-1,...,-N \ ,
\eqn\gvacuumd
$$
with $\mu =0,1,..., D-1$.
To turn this scalar-operator theory into a
${\bf Z}_C$ closed string theory we have to
impose  Virasoro constraints. Analogously to
the continuum theory,
these constraints may arise from to a discrete version
of reparametrization invariance in some underlying original
theory,
but here they will be adopted as an {\it ad hoc} prescription.

The ${\bf Z}_C$-Virasoro operators are defined as follows
$$L_n={1\over 2}\sum _{m=-N}^N \ga_{n-m}\cdot\ga_m \ ,
\ \  \bar L_n={1\over 2}\sum _{m=-N}^N \bar\ga_{n-m}\cdot\bar\ga_m
\ \ \ n=\pm 1,...,\pm N
\eqn\vira
$$
$$
L_0={1\over 2}\ga_0^2 +\sum _{m=1}^N :\ga_{-m}\cdot\ga_m :\ ,\ \
\bar L_0={1\over 2}\bar \ga_0^2 +\sum _{m=1}^N :\bar\ga_{-m}\cdot
\bar\ga_m :\ ,
\eqn\lcero
$$
$$
\ga_0^\mu=\bar\ga_0^\mu={1\over 2}l\hat p^\mu\ .
$$
Normal ordering is defined as usual by placing annihilation
operators to the right.

Throughout this work the constant
$l^2=2\alpha '$, if not explicitly displayed, is assumed
to be 1.

As we shall see below, the operators $L_n$ and $\bar L_n$ generate
a discrete remnant of conformal transformations .
In the continuum limit they become
the Fourier modes of the components $T_{--}$ and $T_{++}$ of the
energy-momentum tensor,
whose discrete analogs are
$$T_{--}=l^2 \sum _{n=-N}^N L_n \omega^{-nx^-}\ ,
\ \ \ \bar T_{++}=l^2 \sum _{n=-N}^N \bar L_n \omega^{-nx^+}\  \ .
\eqn\tensor
$$

The physical space is defined to be the space generated by those states
in the Fock space satisfying
$$
L_n|{\rm phys}\rangle =\bar L_n|{\rm phys}\rangle =0\ ,\ \ \ n=1,2,...,N \ \ ,
\eqn\condition
$$
$$
L_0|{\rm phys}\rangle =\bar L_0|{\rm phys}\rangle =a|{\rm phys}\rangle\ \ ,
\eqn\lcerocon
$$
where $a$  is a constant which will be studied later.

It is remarkable that these operators $L_n, \bar L_n$,
which {\it prima facie} look like a naive truncation
of the usual Virasoro operators, form a closed algebra.
Here the word `algebra' is used in a generalized sense.
Because indices are `angular' variables, i.e.
$n={\bf Z}\ {\rm Mod}\ (2N+1)$ the structure constant
in eq. \modesc \ is not well-defined. Indices must be replaced
by equivalent classes, $[n]\equiv n\ {\rm Mod}\ 2N+1$.
This characteristic arises in many quantum mechanical systems,
for example, a point particle on a circle. The exponentiation
of the algebra usually permits the obtainment of unambiguous results.
Thus the basic commutator
is $[\ga_{[n]}^\mu,\ga _{[m]}^\nu ]=[n]\delta_{[n+m]}\eta^{\mu\nu}$.
For clarity in the formulas we will often omit the brackets of
the indices. We proceed as in the usual operator formalism of the bosonic
string theory,  but a little bit of more care is necessary.
Let us first calculate $[L_m,L_n]$ with
$m\neq -n$ so we do not need to keep track
of the anomalous terms.
The anomaly will be considered in the next section.
$$
\eqalign {[L_m,L_n]=&{1\over 4}\sum_{k,l=-N}^N [\ga_{m-k}\cdot
\ga_k,\ga_{n-l}\cdot\ga_l ]\cr
=&{1\over 4}\sum_{k,l=-N}^N \big( k\ga_{m-k}\cdot\ga_l \delta_{k+n-l} +
(m-k)\ga_k\cdot\ga_l \delta_{m+n-k-l}\cr
& +k\ga_{m-k}\cdot\ga_{n-l} \delta_{k+l}+(m-k)\ga_{k}\cdot\ga_{n-l}
\delta_{m-k+l} \big) \ .\cr}
\eqn\elemnu
$$

\noindent  Now we perform the sum over $l$, bearing in mind that
$\delta_{[\pm (2N+1)]}=1$. We get
$$
[L_m,L_n]={1\over 2}\sum_{k=-N}^N k\ga_{m-k}\cdot\ga_{n+k}
+{1\over 2}\sum_{k=-N}^N (m-k)\ga_k\cdot\ga_{m+n-k} \ \ .
\eqn\elemnd
$$

\noindent Changing variables $k'=k+n$ in the first sum, eq. \elemnd \ becomes
$$
[L_m,L_n]={1\over 2}\sum_{k'=-N+n}^{N+n} (k'-n)\ga_{m+n-k'}\cdot\ga_{k'}
+{1\over 2}\sum_{k=-N}^N (m-k)\ga_k\cdot\ga_{m+n-k} \ .
\eqn\elemnt
$$
Now we can write (e.g. $n>0$)
$$
\sum_{k'=-N+n}^{N+n}=\sum_{k'=-N}^{N}-
\sum_{k'=-N}^{-N+n-1}+\sum_{k'=N+1}^{N+n}\ \ .
$$
The second and
third sums cancel since $[k]=[k+2N+1]$. Therefore the result is

$$
[L_m,L_n]=[m-n]L_{m+n}\ ,\ \ \ m\neq -n \ \ .
\eqn\result
$$

Eq. \result\ resembles the usual Virasoro algebra
with an identification
between generators that produce the same action on ${\bf Z}_C \subset S^1$
(we do not `look' at intermediate points). Strictly speaking, eq. \result\
is not a Lie algebra because the
structure constants are not single-valued;
they must be understood as equivalent classes
\foot {For studies on integrable systems with discrete time in the
mathematical literature see, e.g., ref. \veselov .}.
This will lead to some ambiguities on physical quantities.
In the context of the cellular automaton, this is not a problem:
every physical question can be unambiguously answered in terms
of the primitive variables. The ambiguities arise in trying to
describe the automaton system with the usual tools of quantum mechanics.
In other words, in trying to match life in the cellular automaton with
the continuum physics. Thus the main
problem is the removal of any indefinition in
the continuum limit.

In the continuum theory eq. \result\ is recognized as the
algebra of infinitesimal diffeomorphisms of $S^1$.
Now a complete basis for deformations $x\to x+f_x$
of the discrete circle ${\bf Z}_C$ is provided by the operators

$$
\hat D_n:\ \ \hat D_n F_x =\omega^{nx}\sum_{y=1}^C F_y D_{xy}\ \ ,
\ \ \ \ \ \ D_{xy}\equiv {1\over C} \sum_{n=-N}^N n\omega^{-n(x-y)}\ ,
\eqn\dene
$$
where $F_x$ is  an arbitrary map from the discrete circle
${\bf Z}_C$ to ${\bf R}$. From the definition eq. \dene\  it follows
that $\hat D_n$ satisfies the Leibnitz rule, i.e. given two arbitrary
maps $F_x$ and $G_x$ from ${\bf Z}_C$ to ${\bf R}$ one has the relation
$$
\hat D_n(F_xG_x)=F_x\hat D_nG_x + (\hat D_nF_x) G_x\ \ .
$$
These operators are readily seen to obey the `algebra'
$$
[\hat D_m,\hat D_n]F_x=[m-n]\hat D_{m+n} F_x \ \ ,
\eqn\algden
$$
which is the same as the algebra \result \ of the $L_n\ ,\ \ \bar L_n$.
Introducing $\theta={2\pi \over C}x$, we find
in the continuum $C\to \infty $ limit
$$
\hat D_n \to i e^{in\theta}{\p\over \p \theta}\ ,\ \
\eqn\derivas
$$
i.e. they become the standard basis for diffeomorphisms
of $S^1$.

In the open ${\bf Z}_C$ string theory $\vp^\mu $ will be given by
(cf. eq. \vpopen )

$$
\vp _{x,t}^\mu =\hat \phi^\mu +\beta \hat p^\mu t
+i{l\over 2} \sum_{ n=-N \atop n\neq 0 }^N
{1\over n}\ga_n^\mu (\xi ^{-nx^-}
+\xi ^{-nx^+})\ ,
\eqn\vpopenn
$$
where
$$\xi=e^{i\pi\over N+1}\ ,\ \ \ \ \ \ \ \beta={\pi l^2\over N+1}
\ .
$$

It is easy to show that $\vp^{\mu}_{x,t}$
satisfies the boundary
 condition $\hat D_0 \vp _x^\mu =0$ at the extremal $x=0,N+1$ ,
where $\hat D_0$ is defined in eq. \dene .

The expressions for the $L_n$ , $n=-N,...,N$ are the same as in the
${\bf Z}_C$ closed string case, eqs. \vira , \lcero , but with
$\ga^\mu_0=l\hat p^\mu$.  Note that mode operators $\an ^\mu $  and Virasoro
operators $L_n$ are identified through
$n={\bf Z} \ {\rm Mod} \ (2N+1)$ except that now the
number of cells is $N+2$, unlike
the closed cellular automaton where this relation shows up in a circle with
$2N+1$ or $2N+2$ cells.

\bigskip

\chapter{Central extension}

Let us consider the commutator $[L_n,L_{-n}]$. It has a central term
coming from normal ordering contributions
$$
[L_n,L_{-n}]=[2n]L_0  + A_n\ \ .
\eqn\ano
$$
The explicit computation of the cocycle can be performed
by accounting for normal ordering in eq.\elemnd . Before doing
this, it is worth noting some important differences with
the $N=\infty $ theory. For illustrative purposes we shall
frequently make use of the example provided by the other
extremal, $N=1$.  The ${\bf Z}_C$-Virasoro operators are given by
$$
L_1={1\over 2}\ga_1\cdot\ga_0 +{1\over 2}\ga_0\cdot\ga_1 +{1\over 2}
\ga_{2}\cdot\ga_{-1}=\ga_1\cdot\ga_0 +{1\over 2}\ga_{-1}\cdot\ga_{-1}\ \ ,
\eqn\luno
$$
$$
L_0={1\over 2}\ga_0^2+\ga_1\cdot\ga_{-1}\ \ ,
\eqn\lcero
$$
$$
L_{-1}=\ga_{-1}\cdot\ga_0 +{1\over 2}\ga_{1}\cdot\ga_{1}\ \ .
\eqn\lmuno
$$
Note the unusual terms ${1\over 2} \ga_{\pm 1}\cdot
\ga_{\pm 1}$ in $L_{\mp 1}$
(in the case of arbitrary $N$ the corresponding terms in
$L_n ,\  n=1,...,N$, are  ${1\over 2}
\sum_{m=-N}^{-N+n-1} \ga_{n-m}\cdot\ga_m $, and similarly
for $n=-1,...,-N$).
As a result $L_1|0\rangle\neq 0\ ,\ \ \langle 0|L_{-1}\neq 0$.
This is related to the absence of a M\" obius residual symmetry
in this discrete theory (for a further discussion see sect. 8).
{}From eqs. \luno, \lmuno\ one obtains
$$
\eqalign{ [L_1,L_{-1}]&=\ga^2_0 -\ga_{-1}\cdot\ga_{1}-{D\over 2} \cr
&=\ga^2_0+ 2\ga_{-1}\cdot \ga_{1} -{D\over 2}\cr
&=2L_0 -{D\over 2}\ \ .\cr}
\eqn\lulm
$$
A central term, absent in the corresponding commutator of the continuum
string theory, has appeared. To understand its nature, we now
consider the general $N$ case. From eq. \elemnd \ we obtain ($0<n\leq N$)
$$
[L_n,L_{-n}]={1\over 2} \sum_{m=-N}^N (m+n):\ga_{-m}\cdot\ga_m:
-{D\over 2}\sum_{m=-N}^{-1} m(m+n)
$$
$$
-{1\over 2}\sum_{m=-N}^N
m:\ga_{-n-m}\ga_{n+m}: +{D\over 2}\sum_{m=-N}^{-n-1} m(m+n)
+{D\over 2}\sum_{m=N-n+1}^{N} m(m+n)
\eqn\anom
$$
Hence
$$
\eqalign {A_n&=-{D\over 2}\sum_{m=-n}^{-1}m(m+n)
+{D\over 2}\sum_{m'=1}^{n}(m'+N)(m'+N-n)\cr
&={D\over 2}n N(N+1) \ .\cr }
\eqn\anoma
$$
This is a trivial cocycle which can be removed by redefining
$$L_0\to L_0+{D\over 2} {N(N+1)\over 2}\ .$$
 In fact, this is precisely the number
which arises upon normal ordering of $L_0$:
$$
{1\over 2}\sum_{m=-N}^N\ga_{-m}\cdot\ga_m=
{1\over 2}\sum_{m=-N}^N:\ga_{-m}\cdot\ga_m:+{D\over 2} {N(N+1)\over 2}\ .
\eqn\normal
$$

Thus the theory is free of anomalies for any value of $D$. This may be
no surprise, since anomalies are related to short-distance behaviour
of correlators and here correlations functions are well behaved and
finite at coincident points.
We will find a confirmation of the anomaly freedom in the next section
when we study the Lorentz algebra in the light-cone gauge.

Another way to obtain $A_n$ is by calculating
$\langle 0|[L_n,L_{-n}]|0\rangle $ explicitly using the algebra
of mode operators and the definition of the vacuum given in
eqs. \gvacuum , \gvacuumd  .
One easily finds ($0<n\leq N$)
$$
\langle 0|L_n L_{-n}|0\rangle ={D\over 2}\sum_{m=1}^{n-1}
m(n-m) = {D\over 2}{n^3-n\over 6}
\eqn\consis
$$
 where the well-known formulas
$\sum_{m=1}^n m={1\over 2} n(n+1)\ ,\ \ \sum_{m=1}^n m^2=
{1\over 3}n^3+{1\over 2}n^2+{1\over 6}n  $, have been employed.
On the other hand, observing that
$$L_n|0\rangle ={1\over 2}
\sum_{m=-N}^{-N+n-1}\ga_{n-m}\cdot \ga_m |0\rangle \neq 0\ ,
$$
we get
$$
\langle 0|L_{-n} L_{n}|0\rangle = {D\over 2}{n^3-n\over 6}
-{D\over 2} n N(N+1)\ \ ,
\eqn\consist
$$
i.e. $\langle 0|[L_n,L_{-n}]|0\rangle ={D\over 2} n N(N+1)$,
in accordance with eq.\anoma .

\bigskip

\chapter{ Light-Cone Gauge }

The residual symmetry generated by the
$L_n, \bar L_n$ can be used to gauge away all the oscillator modes
of one `target' coordinate. Indeed, these operators
generate the following transformations on the operators
$\vp^{R\mu} ,\  \vp^{L\mu}$
$$
[L_n,\vp^{R\mu}_x]=-{i\over 2}\sum_{m=-N}^N \ga_{n+m}^\mu\ \omega^{mx^-}
\ ,\ \ \ \
[\bar L_n,\vp^{L\mu}_x]=-{i\over 2}\sum_{m=-N}^N \bar \ga_{n+m}^\mu\
\omega^{mx^+}\ .
\eqn\uno
$$
Thus we use this residual symmetry to set
$$
\vp^+=\phi^+ + {\pi l^2\over C}p^+t\ ,\ \ \ \ \ \
\vp^\pm={\vp^0\pm \vp^{D-1}\over \sqrt {2}}\ \ .
\eqn\lcg
$$

Eq. \lcg \  means, in particular, that in this physical
gauge the target time $\vp^+ $ has in some sense
a discrete structure.
As a result, the target
space light-cone energy $p_- $ will be {\it periodic},  i.e.
$$
p_- \cong p_- + {4C\over l^2p^+} {\bf Z}\ .
\eqn\energia
$$
It is natural to require that all
the target-space coordinates have this type of discrete structure,
though
we will not explore this possibility here.
The consequences and interpretation of eq. \energia\
will be considered later.

Since right and left moving sectors play a symmetrical role,
it is sufficient to consider the open string case.
The ${\bf Z}_C$ Virasoro constraint equations $L_0~=~a$, \
$L_n~=~0~, \ \ n~=~\pm 1,...,\pm N$
can be solved giving
$$
l_n\equiv p^+\an^-={1\over 2}\sum_{i=1}^{D-2}\sum_{m=-N}^N
:\ga_{n-m}^i\ga_m^i: -a \delta_n\ \ .
\eqn\menos
$$
It is straightforward to show that the $l_n $ satisfy the
algebra
$$
[l_m,l_n]=[m-n]l_{m+n}+\delta_{m+n}\big[{D-2\over 2}m N(N+1)
+2am\big] \ .
\eqn\lcga
$$
The calculation is analogous to the discussions in sects. 2, 3.

While the light-cone formalism is manifestly free of negative-norm states,
it is not manifestly covariant.
The question is whether the theory is really Lorentz
invariant in this gauge. In the standard bosonic string theory,
it turns out that only if $D=26$ and $a=1$ the Lorentz algebra
is anomaly free (for a review see ref. [\gsw ]).
For other values of $D$ and $a$ it is not possible
to fix the light-cone gauge maintaining at the same time
the Lorentz covariance of the theory.

The Lorentz generators $J^{\mu\nu} $ are
$$ J^{\mu\nu}=l^{\mu\nu}+E^{\mu\nu} \ ,$$
$$
l^{\mu\nu}=\phi^\mu p^\nu-\phi^\nu p^\mu \ ,\ \ \
E^{\mu\nu}=-i\sum _{n=1}^N {1\over n}(\ga^\mu_{-n}\an^\nu -
\ga_{-n}^\nu\an^\mu) \ .
\eqn\lorentz
$$
It is therefore important to verify that these operators really generate
the Lorentz algebra. Most of the commutators can be carried out in a
straightforward way and they give the correct answer. The potential anomaly
comes from $[J^{i-},J^{j-}]$ --which must vanish if Lorentz invariance
is to hold-- since the transformations generated by $J^{i-}$ affect
the light-cone gauge choice.

By using eq. \lorentz\  and the algebra of the oscillator mode operators
it is easy to prove that
$$
[J^{i-},J^{j-}]=-(p^+)^{-2} C^{ij}
\eqn\cij
$$
with
$$
C^{ij}=2il_0E^{ij}-[E^i,E^j]-iE^ip^j+iE^jp^i\ ,\
\ \ \ \ E^i\equiv p^+E^{i-}\ .
\eqn\ccij
$$
Let us first verify that the classical part vanishes.
For the moment we will drop the potential anomalous terms, i.e. all
terms quadratic in oscillators. They will be considered in detail
later. Using eq. \lcga \ and
$$
[\ga_m^i,l_n]=m\ga_{m+n}^i \ ,
\eqn\amln
$$
we find
$$
[E^i,E^j]={\cal A}+{\cal B}+{\cal C}\ ,
\eqn\masaun
$$
with
$$
{\cal A}=-\sum_{n,m=1}^N{1\over n} \ga^i_{-n}\big( \ga_{n-m}^jl_m -
\ga_{-m}^jl_{n+m}\big) - (i\leftrightarrow j)\ \ ,
\eqn\aaaa
$$
$$
{\cal B}=\sum_{n,m=1}^N{1\over n} \big( l_{-m}\ga_{m-n}^j -
l_{-n-m}\ga_{m}^j\big)\ga^i_{n} - (i\leftrightarrow j)\ \ ,
\eqn\bbbb
$$
$$
{\cal C}=\sum_{n,m=1}^N \big({1\over n} + {1\over m} \big)
\ga_{-n}^il_{n-m}\ga^j_m- {1\over n} \ga_{-n}^il_{-m}\ga^j_{n+m}
-{1\over m} \ga_{-n-m}^i\ga^j_{m}l_{n}
- (i\leftrightarrow j)\ \ .
\eqn\cccc
$$
In the above eqs. \aaaa, \bbbb \ and  \cccc \  some terms cancel out
and what remains is
$$
{\cal A}=-\sum_{n=1}^N \big( \sum_{m=1}^n-\sum_{m=N+1}^{N+n}\big)
{1\over n} \ga^i_{-n}\ga_{n-m}^jl_m
- (i\leftrightarrow j)\ \ ,
\eqn\aaaaa
$$
$$
{\cal B}=\sum_{n=1}^N\big( \sum_{m=1}^n - \sum_{m=N+1}^{N+n}\big)
{1\over n}  l_{-m}\ga_{m-n}^j \ga^i_{n} - (i\leftrightarrow j)\ \ ,
\eqn\bbbbb
$$
$$
{\cal C}=\sum_{m=1}^N \big(\sum_{n=1-m}^0 -\sum_{n=N-m+1}^N
\big){1\over m} \ga_{-n-m}^il_{n}\ga^j_m
$$
$$+\sum_{n=1}^N \big(\sum_{m=1-n}^0 -\sum_{m=N-n+1}^N
\big){1\over n} \ga_{-n}^i l_{-m}\ga^j_{m+n}
- (i\leftrightarrow j)\ \ .
\eqn\ccccc
$$
The second sum in the parentheses is absent in the $N=\infty$
theory. By using the fact $n = n\ {\rm Mod}\ {2N+1}$ and conveniently
renaming variables
${\cal C}$ can be written in the following form
$$
{\cal C}=\sum_{n=1}^N \big(\sum_{m=1}^n -\sum_{m=N+1}^{N+n}
\big){1\over n} \ga_{n-m}^il_{m}\ga^j_{-n}
+\sum_{n=1}^N\big( \sum_{m=1}^n - \sum_{m=N+1}^{N+n}\big)
{1\over n}  l_{-m}\ga_{m-n}^i \ga^j_{n}
$$
$$
+2\sum_{m=1}^N{1\over m} \ga^i_{-m}\ga^j_m l_0
-\sum_{n=1}^N{1\over n}\ga_0^i (l_{-n}\ga_n^j -l_{n}\ga_{-n}^j )
- (i\leftrightarrow j)\ \ .
\eqn\manyc
$$
Therefore
$$
{\cal C}=-{\cal A}-{\cal B}
+2il_0E^{ij}-iE^ip^j+iE^jp^i\ \ .
\eqn\morec
$$
Thus, comparing with eqs. \cij , \ccij  \ and \masaun \  we see that
$[J^{i-},J^{j-}]|_{\rm classical}=0$.

Let us now consider the computation
of the anomaly. We follow the analogous
derivation given in ref. [\gsw ].
Since the classical part is zero,
$C^{ij}$ can only be quadratic in the oscillator modes. In fact it
can only be of the form
$$
C^{ij}=\sum_{m=1}^N\Delta_m (\ga_{-m}^i\ga_m^{j}-\ga_{-m}^j\ga_m^{i})
\ \ .
\eqn\cccij
$$
The coefficients $\Delta_m$ can be determined by evaluating
the matrix elements
$$
\langle 0|\ga^k_m C^{ij}\ga_{-m}^l|0\rangle =m^2(\delta^{ik}\delta^{jl}
-\delta^{jk}\delta^{il})\Delta_m\ \ .
\eqn\deltam
$$
By explicit computation we find
$$
\langle 0|\ga_m^k( -iE^ip^j +i E^j p^i )\ga^l_{-m}
|0\rangle =-\delta^{ik}p^jp^l m
+\delta^{il}p^jp^k m -(i\leftrightarrow j)\ \ ,
\eqn\primo
$$
and
$$
2i \langle 0|\ga^k_m l_0 E^{ij}\ga^l_{-m}
|0\rangle =m(p^2_i-2a+2m)(\delta^{jl}\delta^{ki}
-\delta^{jk}\delta^{il})\ \ .
\eqn\secondo
$$

The calculation of $\langle 0|\ga_{m}^k [E^i,E^j]\ga^l_{-m} |0\rangle $
is more involved than the evaluation of the corresponding matrix element
in the continuum theory because now $l_n$ and $l_{-n}$
do not destroy $|0\rangle $ and $\langle 0|$. Let us write
$$
\langle 0|\ga_{m}^k [E^i,E^j]\ga^l_{-m} |0\rangle =A_{\rm I}+A_{\rm II}
+A_{\rm III}
\eqn\aaa
$$
where
$$
\eqalign {A_{\rm I}&=\langle 0|l_m l_{-m} |0\rangle (\delta^{jl}\delta^{ki}
-\delta^{jk}\delta^{il}) \ ,\cr
A_{\rm II}&=-\sum_{n=1}^m \big(\delta^{jl}\langle 0|\ga^k_m l_{-n}\ga^i_
{n-m}|0 \rangle + \delta^{ki}\langle 0|\ga^j_{m-n} l_{n}\ga^l_{-m}
|0 \rangle  -(i\leftrightarrow j) \big)\ ,\cr
A_{\rm III}&=\sum_{n,n'=1}^N {1\over nn'}
\langle 0|\ga^k_m l_{-n}\ga^i_n\ga^j_{-n'}l_{n'}\ga_{-m}^l |0 \rangle \ .\cr}
\eqn\quiaaa
$$
We find
$$
A_{\rm II}=-\delta^{jl}\delta^{ki} m^2(m-1) -m\delta^{ki} p^jp^l
-m\delta^{jl}p^kp^i -(i\leftrightarrow j)\ .
\eqn\hados
$$
By straightforward algebra we can take $A_{\rm III}$ to
the form
$$
A_{\rm III}=\delta^{il}\delta^{kj}\langle 0|l_{-m}l_m|0\rangle
+\delta^{il}\sum_{n'=1}^N\langle 0|\ga^k_m \ga^j_{-m-n'}l_{n'}|0\rangle
+\delta^{kj}\sum_{n=1}^N\langle 0|l_{-n}\ga^i_{m+n} \ga^l_{-m}|0\rangle
$$
$$+ \sum_{n,n'=1}^N \langle 0|\ga_m^k\ga^j_{-n-n'}\ga^i_{n+n'}\ga^l_{-m}
|0\rangle -(i\leftrightarrow j)\ \ ,
\eqn\atres
$$
obtaining
$$
A_{\rm III}=(m^3+m^2-\langle 0|l_{-m}l_m|0\rangle )
(\delta^{jl}\delta^{ki}-\delta^{il}\delta^{kj})\ \ .
\eqn\atresf
$$
Using eqs. \quiaaa, \hados , \atresf \  and the commutation relation
\lcga\ we arrive at the following expression:
$$
\langle 0|\ga_{m}^k [E^i,E^j]\ga^l_{-m} |0\rangle =
m\big( {D-2\over 2}N(N+1)+2m +p^2_i\big)
(\delta^{jl}\delta^{ki}-\delta^{il}\delta^{kj})\ \
$$
$$
-m\big(\delta^{ki}p^jp^l+\delta^{jl}p^kp^i-
(i\leftrightarrow j)\big)\ \ .
\eqn\eiej
$$
Inserting eqs. \primo, \secondo, \eiej \ into eq. \deltam \
we find (see also eq. \ccij )
$$
\Delta_m=-{1\over m}\big( {D-2\over 2} N(N+1) +2a \big)\ \ .
\eqn\fin
$$
Requiring $\Delta_m=0 $ gives
$$
a=-{D-2\over 2}{N(N+1)\over 2}\ \ ,
\eqn\siah
$$
with no additional restriction on $D$.

The value of $a$ is just the same as the value of the constant
which emerges in normal ordering
the operator $l_0$ and it is in agreement with the value obtained
in the usual $N= \infty $ case if one makes use of
zeta-function regularization to compute the infinite sum [\gsw ],
i.e.
$\zeta (s)=\sum_{n=1}^\infty n^{-s}$, with $\zeta (-1)=-{1\over 12}$,
thus giving
$$a={D-2\over 24}\ ,$$
which yields $a=1$ for $D=26$. The correspondence with the standard
bosonic string theory in the continuum limit is more carefully examined
in sect. 8.

\bigskip

\chapter {Spectrum}

Along this section we will continue our investigation on
properties inherent to
the discrete theory, leaving the discussion on the approximation
to the continuum theory  deferred to sect.~8.

The physical states are much more easily constructed in the light-cone
gauge, especially in this discrete theory, since
in the covariant formalism physical states
will involve states from different levels
(see below).
The well-known disadvantage of the light-cone gauge is that
the states appear as multiplets of $SO(D-2)$, the transverse rotation
group, even though the proof of Lorentz invariance guarantees that the
massive levels fill out complete multiplets of $SO(D-1)$.
A general physical state in the closed ${\bf Z}_C$ string theory
can be written in the form
$$
|{\rm phys}\rangle = \xi _{\{ i,\bar i\} }(p)
\prod_{n=1}^N (\ga_{-n}^{i_1} ...\ga_{-n}^{i_{\epsilon _n}})
(\bar\ga_{-n}^{\bar i_1}...\bar\ga_{-n}^{\bar i_{\bar\epsilon_n}} )
|0;p \rangle \ \ ,
\eqn\fisico
$$
where $\xi _{\{ i,\bar i\} }$ is an arbitrary polarization vector and
the $\epsilon_n, \bar \epsilon_n $ are subject to the constraint
$$
\sum_{n=1}^N n\epsilon_n =\sum_{n=1}^N n\bar\epsilon_n  \ \ ,
$$
coming from the $l_0 =\bar l_0 $ condition.

The mass spectrum follows from eq. \menos :
$$
{1\over 8}l^2m^2=-a+\sum_{n=1}^N \ga_{-n}^i\an ^i=
-a + \sum_{n=1}^N \bar \ga_{-n}^i \bar \an ^i \ .
\eqn\masas
$$
or
$$
{1\over 8}l^2m^2=-a +\sum_{n=1}^N [n] a_{n}^{\dag i} a_n^i
\eqn\masses
$$

The above equations imply that $m^2$ is also an `angular'
or periodic variable, i.e.
$$
{1\over 8}l^2m^2 \cong {1\over 8}l^2m^2 +kC\ ,\ \ \ k={\rm integer},
\eqn\drama
$$
This indefinition is also inferred from  eq. \energia .

The fact that time discretization leads to identifications
of the form energy$\cong $ energy $+2\pi n$ is not new. The novelty here
is that this property, which is of course present in the world-sheet
theory, is also reflected by eq. \drama \ to the target space.
In particular, eq. \drama\
implies that in this theory the masses of
${\bf Z}_C$-string excitations of arbitrary spin
can be reduced to the interval
${l^2\over 8} m^2=-a,-a+1,...,-a+C-1$.

According to eq. \masas , the mass of the state \fisico \ will be
given by
$$
{1\over 8}l^2 m^2= -a+\sum_{n=1}^N n\epsilon_n \ +{\bf Z}C\ .
\eqn\mas
$$
Thus the states with $\sum_{n=1}^N n\epsilon_n \geq C$ will repeat
the same mass pattern. There is no analog of this degeneracy
in the continuum
string theory.
For example, let $\epsilon_n=C\delta_{n-1} $\ ; this higher-level
state has mass
${1\over 8}l^2 m^2= -a+C=-a$, i.e. the same square mass
as the `tachyon' state. For each ${l^2\over 8} m^2=-a,-a+1,...,-a+C-1$,
there will be an infinite tower of higher-spin
states with the same square mass.

In the covariant formalism the expressions for physical states
are complicated, even at the lowest level. Establishing a one-to-one
correspondence with light-cone gauge spectrum is important, since
this is the basis of
the proof of the no-ghost theorem, this gauge being manifestly
ghost free.

The mass of a given physical state follows from the
mass-shell condition
$L_0|{\rm phys}\rangle =\bar L_0|{\rm phys}\rangle =a|{\rm phys}\rangle\ $\ ,
$$
{1\over 8}l^2m^2=-a+\sum_{n=1}^N \ga_{-n}\cdot \an =
-a + \sum_{n=1}^N \bar \ga_{-n}\cdot \bar \an \ \ ,
\eqn\masasa
$$
which is the same as the mass-shell condition one finds in the
light-cone gauge treatment, except that now all
oscillators contribute to $m^2$.

Let us consider the $N=1$ example in the case of open strings.
The entire spectrum of $N=1$ open
string consists of one scalar field $T$ of mass $m^2_T=-2a$, one vector field
$A_\mu$ of mass $m^2_A=-2a+2$, one second-rank tensor $B_{\mu\nu}$ of mass
$m^2_B=-2a+4$,
plus  towers of states
of masses $-2a+{\bf Z}C=m^2_T$, $-2a+2+{\bf Z}C=m^2_A$, and
$-2a+4+{\bf Z}C=m^2_B$.
An interesting problem is deriving the effective field theory action for
these states. The linearized equations of motion for all fields
are easily obtained from the Virasoro condition $L_1|{\rm phys}
\rangle =0$.

Let us explicitly derive the `tachyon' state.
Using eqs. \luno , \lcero\  we see that

$$
L_0|0;p^\mu \rangle ={1\over 2} p^2 |0;p^\mu \rangle\ ,\ \ \
L_1 |0;p^\mu \rangle ={1\over 2}\ga_{-1}\cdot \ga_{-1}|0;p^\mu
\rangle \neq 0\ \ .
\eqn\ones
$$

\n Therefore we find, somewhat unexpectedly, that
the state $|0;p^\mu \rangle $ is not physical. This is
 unlike its continuum counterpart. Moreover,
how can we explain the discrepancy
with the light-cone gauge?. In order to clarify this controversy,
let us find the correct `tachyon' physical state. From the
structure of $L_1$ we see that the `tachyon' state must be
of the form

$$
|T (p^\mu )
\rangle =(1+ A^{(1)}_{\mu_1 \mu_2 \mu_3}\ga_{-1}^{\mu_1}
\ga_{-1}^{\mu_1}\ga_{-1}^{\mu_1} +
A^{(2)}_{\mu_1 ...\mu_6}\ga_{-1}^{\mu_1}...\ga_{-1}^{\mu_6}
+...) |0; p^\mu \rangle  \ \ ,
\eqn\tt
$$

\n i.e. a linear combination
involving all states with $p^2=p^2+{\bf Z}C,\ C=3$.
The physical condition $L_1|T(p^\mu )\rangle =0$ is solved with

$$
A^{(1)}_{ \mu_1\mu_2 \mu_3}=-{1\over 6 p^2}
(p_{\mu_3}\eta_{\mu_1 \mu_2}  +
p_{\mu_2}\eta _{\mu_1 \mu_3 }+p_{\mu_1}\eta_{\mu_2 \mu_3 })
+{1\over 3 p^4}p_{\mu_1 }p_{\mu_2 }p _{\mu_3} \ \ ,
\eqn\soluno
$$
etc. . We can also
check that the state \tt \ satisfies the $L_0$ condition

$$
\eqalign{ L_0 |T (p^\mu )\rangle &= {1\over 2} (p^2 +{\bf Z}C)
|T (p^\mu )\rangle + \big( [3] A^{(1)}_{\mu_1 \mu_2 \mu_3}\ga_{-1}^{\mu_1}
\ga_{-1}^{\mu_2}\ga_{-1}^{\mu_3} + [6]
A^{(2)}_{\mu_1 ...\mu_6}\ga_{-1}^{\mu_1}...\ga_{-1}^{\mu_6}
+...\big)|0; p^\mu \rangle   \cr
&= {1\over 2} p^2 |T(p^\mu )\rangle  \cr}
\eqn\cerapio
$$
Now we understand what is happenning: in the covariant
formalism the `tachyon' physical state $|T (p^\mu )\rangle $
contains, beside
$|0;p^\mu \rangle $, a linear combination of all {\it longitudinal}
components of the higher-spin
states with the same mass $p^2=-2a$. In this $N=1$ example
they are in fact those having $kC$ oscillator modes $\ga_{-1}$,
with $k=1,2,...,\infty $.
In the general case the lowest-level
physical state will have the form
$$
|T(p^\mu )\rangle =\sum_{r=0}^\infty \sum_{n_1,...,n_r=-N}^N
\delta_{[n_1+...+n_r]}
A_{\mu_1 ...\mu_r}\ga_{-{n_1}}^{\mu_1}...\ga_{-{n_r}}^{\mu_r}
|0; p^\mu \rangle
\eqn\tachgen
$$
where $A_{\mu_1 ...\mu_r}$ are longitudinal components of the
states. In the light-cone gauge, clearly,
these longitudinal components are automatically gauged away
and $|T(p^\mu )\rangle $ simply reduces to $|0;p^\mu \rangle $.
Correspondence between the light-cone
gauge and the covariant formalism requires that
these longitudinal states (which, just as in standard
string theory, can be expressed as
$\sum_{n=1}^N L_{-n}|\chi \rangle $)
decouple from scattering amplitudes.

\bigskip

\chapter {Vertex Operators}

In standard string theory conformal invariance permits the mapping
of an arbitrary genus Riemann surface with $p$ tubes extending into the
far past and the far future --corresponding to incoming and outgoing
strings-- to a Riemann surface of the same topology and moduli
with $p$ punctures replacing the tubes, the quantum numbers
of the external strings being represented by vertex operators at
each puncture. Since the discrete theory is not conformal invariant,
in principle it is not clear than
an arbitrary external state can be replaced by
a vertex operator. Nevertheless, in order to make contact with the continuum
theory, it is useful to introduce the discrete
analog of these operators.

In the bosonic closed string theory
a vertex operator $W$ for an on-shell string state
is a local operator of conformal dimension (1,1), i.e. satisfying
$$
\eqalign {
[L_n, W(z,\bar z)]&=z^{n+1}{\p\over\p z} W +nz^n W\ ,\ \ \ \cr
[\bar L_n,W(z,\bar z)]&=\bar z^{n+1}{\p\over\p \bar z} W +n
\bar z^n W\ ,\ \ \ \cr }
\eqn\oneone
$$
which carries the correct Lorentz quantum numbers of the
represented particle.

In analogous way we will say that a local operator $W_{x,t}$
has ${\bf Z}_C$-{\it conformal dimension}
$(\lambda , \lambda )$ if it obeys the following commutation relations
$$
[L_n, W_{x^+,x^-}]=-D_n^- W+\lambda n \omega ^{nx^-}W\ ,
\eqn\wwr
$$
$$
[\bar L_n, W_{x^+,x^-}]=-D_n^+ W+\lambda n  \omega ^{nx^+}W\ ,
\eqn\wwl
$$
where $D_n^-,\ D_n^+$ are operators defined
as in eq. \dene\ , with respect to $x^-,\ x^+$ respectively, and
$n=-N,...,N$.
Similarly, in the case of the open ${\bf Z}_C$ string theory
an operator $W_t$ to be inserted on the boundary is said to have
${\bf Z}_C$-conformal dimension $\lambda $
if it satisfies
$$
[L_n, W_{t}]=-D_n^t W_t+\lambda n \xi ^{nt}W_t\ ,
\ \ \ \ n=-N,...,N \ .
\eqn\wwopen
$$
where
$$
D_n^t: \ \ \ D_n^t F_t=\xi ^{nt}\sum_{t'=0}^{C}
F_{t'}D_{tt'}\ ,\ \ \ \ D_{tt'}={1\over C}\sum_{n=-N}^N
n\xi ^{-n(t-t')}\ .
$$

A consequence of the commutation relations \oneone\ is that vertex
operators can be used to map physical states to
physical states, up to spurious states which decouple from
scattering amplitudes. This property holds true also in the present
case provided the local operator has ${\bf Z}_C$-conformal dimension
$\lambda =1$. Indeed, for $\lambda=1$ one has
$[L_n, W_{t}]=-D_0^t \big( \xi ^{nt}W_t \big) \ $, etc.

The candidate for a vertex operator representing the tachyon
is
$$
W_T(p^\mu)=:e^{ip\cdot\vp}:\ \ \ \ .
\eqn\tachver
$$
This operator injects momentum $p^\mu$ on a single cell $x,t$.
Let us compute the commutator $[L_n, W_T]$. For simplicity we
restrain our attention to the open string case.
The calculation is similar to the corresponding one of the
standard continuum theory, so we will not reproduce all steps.
By using the basic commutator
$$
[L_m,e^{p\cdot \ga_{-n}}]={1\over 2}n (p\cdot\ga_{m-n}
e^{p\cdot \ga_{-n}} +e^{p\cdot \ga_{-n}}p\cdot\ga_{m-n})\ \ ,
\eqn\bascom
$$
one attains a result of the form \wwopen\ with $\lambda=0$
where $W_T$ is not yet normal ordered. The usual anomalous
dimension precisely arises in the normal-ordering process.
In doing this, we get an additional contribution relative
to the continuum case:
$$
\eqalign {
[L_m, W_{t}^T]&=-D_m^t W_t^T+{p^2\over 2}  \xi ^{nt}W^T_t
\big(\sum _{n=1}^m 1 -\sum _{n=-N }^{-N+m-1} 1 \big)
 \cr
&=-D_m^t W_t^T \ \ , \cr}
\eqn\sinanom
$$
that is, the operator \tachver\ has no anomalous dimension, fact to
be expected in view of the absence of short-distance
singularities. The ${\bf Z}_C $-conformal dimension of the operator
$:e^{ip\cdot \vp}:$ is simply zero, independent of $p^2$.

Higher-level vertex operators for the closed ${\bf Z}_C$ string
are obtained in the standard way by successive multiplication
of $e^{ip\cdot\vp}$ by
$$
{i\over 4}l^2\hat p^\mu +D_0^-\vp ^{R\mu} \equiv D_0^{-}\vp^\mu \ ,\ \ \ \ \
{i\over 4}l^2\hat p^\mu +D_0^+\vp ^{L\mu} \equiv D_0^+\vp^\mu
\eqn\nonso
$$
The analogous construction for
higher-level vertices applies to the open ${\bf Z}_C$ string case.
The action of the operator $D_0$ is defined on maps which are
single-valued on ${\bf Z}_C$. The $x^\pm$ are not periodic so in
the identification \nonso\ the symbolic notation
$D^\pm_0 x^\pm =iC/ 2\pi $ is understood (cf. eq. \derivas ).
In addition,
`higher derivative' operators $(D_0^{-})^n\vp^\mu \ ,\ \ (D_0^+)^n
\vp^\mu $ may be present.
For example, the graviton vertex operator is
given by
$$
V_{\rm g}=\xi_{\mu\nu}(p)
D_0^-\vp^\mu D_0^+\vp^\nu e^{ip\cdot\vp}\ ,
\eqn\grav
$$
where ${1\over 8}p^2=-a+1 $ and
$\xi_{\mu\nu}$ is a symmetric tensor obeying the conditions
 $  \xi_{\mu\nu}p^\mu=0 , \ \xi_{\mu}^\mu=0 $\
 which avert the appearance of unwanted
additional terms in eqs. \wwr\ and \wwl .
This vertex has ${\bf Z}_C $-conformal dimension (1,1) and thus
it can be used to map physical states to physical states.
Vertex operators corresponding to levels higher than two
will inevitably have ${\bf Z}_C $-conformal dimension
greater than (1,1).

\bigskip

\chapter {Continuum Limit and Tree-Level Scattering Amplitudes}

The theory we were studying hitherto is different from the bosonic
string theory in a number of important respects.
In this ${\bf Z}_C$-string theory, free of short-distance singularities,
 no anomalous terms have appeared in the algebras.
The computational
reason behind this is the contribution of excitations $\an ^\mu$
with $|n|\sim N$ which has the opposite sign and leads to
exact cancellation of all the anomalies. Indeed,
the well-known conformal anomaly corresponding to
$D$ scalar fields
$$
{D\over 12}(n^3-n)=-{D\over 2}\sum_{m=-n}^{-1}m(m+n)
\ ,
$$
is present in eq. \anom \ but is cancelled
by the term
$$
{D\over 2}\sum_{m=N-n+1}^N m(m+n) \ \ ,
\eqn\esel
$$
which originated from commutators involving $\ga_n^\mu $ with $|n|\sim N$.
Similarly, the usual anomalous terms in the
Lorentz algebra are cancelled from contributions of the type \esel \ and
the same fate undergoes the
usual anomalous dimension for the operator $e^{ip\cdot\vp }$
(cf. eq. \sinanom ).

A similar situation emerged in ref. [\hooftt ]
in the context
of the Fermi one-dimensional shift automaton.
The Schwinger term in the fermionic current algebra is lacking
in the finite $N$ theory.
In the continuum limit, $(x,t) \to (x/d, t/d)$, $C\to\infty , \ d\to 0$,
$Cd=\pi $, one would like to retain only low-energy
excitations of the fermions. By the standard $\epsilon $-prescription
for distributions the contributions from excitations near $\pm N$ are
dampened by a factor $e^{-\epsilon C}$.
The processes of letting $\epsilon \to 0$ and taking
the continuum limit  do not commute. To safely obtain
the continuum theory one has to first take $N\to \infty $ and eventually
take $\epsilon\to 0$.
In doing this all contributions to the anomaly coming from
oscillators with frequencies
$p\sim \pm N$ will vanish as $N\to\infty $.

Another way to eliminate the non-perturbative contribution of
excitations with $n\sim \pm N$
in the continuum limit is by
going to Euclidean world-sheet time, $t\to it=\tau $. The operator
$\vp ^\mu $ corresponding to the closed string case takes the form
$$
\vp ^{\mu}=\hat \phi^\mu +i{\pi l^2\over C}\hat p^\mu \tau
+i{l\over 2}\sum_{n=-N}^N
{1\over n}(\ga_n^\mu\omega ^{-n(i\tau -x)}
+\bar\ga_n^\mu\omega ^{-n(i\tau + x)})\ ,\ \ \ \ \omega \equiv
e^{i{2\pi\over C}}\ \ .
\eqn\truncada
$$
If we are interested
in physics below some (world-sheet) energy scale of order $N$,
then excitations with $|n|\sim N $ will be frozen, and the
`effective' anomalies corresponding to this long-distance theory,
in the absence of terms like e.g. eq. \esel ,
will coincide with the continuum values.

The physical mechanism which would effectively implement
the $\epsilon $-prescription or any other prescription
is unclear to us. We will not attempt any deeper enquiry
on this at this primitive stage, though it is interesting to speculate
on a {\it vinculum} with the Hagedorn phase transition.

In the covariant approach
the critical values $a=1$
and $D=26$ are evidenced by the multiple appearance of
zero-norm states. This holds true also in the present
theory since e.g. the norms of the states
$L_{-1}|0;p^\mu\rangle =\ga_{-1}\cdot p
|0;p^\mu\rangle $, with $p^2=0$,
 and $(L_{-2}+{3\over 2}L_{-1}^2 )|0;p^\mu \rangle $, with $p^2=-2$,
become zero for $a=1 \ {\rm Mod} \ {\bf Z}$ and $D=2(13 \ {\rm Mod} \
{\bf Z})$ respectively.
Thus, in the spirit of making contact with the continuum theory,
let us set $a=1$ and $D=26$ and
define a
tree-level scattering amplitude as a correlator of vertex operators
associated with the external particles which are present in the
scattering process. For scattering of closed ${\bf Z}_C$ string states
$$
A^M_{\rm closed }=\kappa^{M-2}\langle 0|T\{ V^1...V^M\} |0\rangle\ ,
\eqn\amplit
$$
where $\kappa $ is a coupling constant and
$$
V^i=\sum_{t=-\infty}^{\infty }\sum_{x=1}^C \omega^{-bx^-}
\omega^{-bx^+}W^i_{x,t}\ , \ \ \ b={1\over 8}p^2 \ ,
\eqn\vertice
$$
where $W^i_{x,t}
$ are the vertex operators discussed in the previous section.
The scattering amplitudes for open ${\bf Z}_C$ strings are defined in a similar
way ($g^2\sim \kappa $)
$$
A^M_{\rm open}=g^{M-2}\sum_{t_1>...>t_M}
\prod_{i=1}^M \xi ^{-4b_it_i}\langle 0| W_{t_1}^1...W^M_{t_M}
|0\rangle + {\rm cyclic \ perm.}
\eqn\amplito
$$

The main uncertainty in these `scattering amplitudes' is that
only level-one vertices --corresponding to massless particles--
have the correct ${\bf Z}_C$-conformal dimension 1. As discussed
above, only if the effect of excitations with momentum
near $\pm N$ is suppressed an anomalous dimension for the
operator $e^{ip\cdot\vp }$ will arise.
It should be emphasized that here
the scattering amplitudes \amplit \  and  \amplito \
are introduced with no other scope than developing
an approximative method to the continuum theory.

The only remnant of M\" obius symmetry left in the process of
discretization are discrete translations ${\bf Z}_C$. This is not
a surprise, since $SL(2,{\bf C})$ involve rescaling $\delta x=\lambda x$
and $\delta x=c x^2$ which clearly cannot be symmetries in this
${\bf Z}\times {\bf Z}_C$ theory.
As a matter of
fact, the vacuum is not annihilated by $L_{\pm 1}$, $\bar L_{\pm 1}$,
as announced in sect. 4, but it is annihilated by $L_0$, $\bar L_0$
which are the generators of $x^-$ and $x^+$ discrete translations.

By construction scattering amplitudes defined by \amplit\
will approach
the corresponding scattering amplitudes of the continuum theory
as $N\to\infty$, since in this limit the basic
correlator $\langle\vp \vp \rangle$ equals the corresponding correlator
of the standard string theory, and the Wick theorem is applicable.
The only difference is that here the M\" obius symmetry $SL(2,{\bf C})$
(or $SL(2,{\bf R})$ for the open string)
is only an approximate symmetry in low-energy processes which do not
probe the discrete world-sheet structure.
The discrete translation invariance permits to fix the position of
one vertex operator.
For example, setting $x_M, t_M=0$,
the $M$-tachyon amplitude will be given by
$$\eqalign{
A^M_T &=\sum_{t_1>...>t_{M-1}>0}\ \sum_{x_1,...,x_{M-1}=1}^C
\big( \prod_{i=1}^{M-1}\omega^{-2b_it_i}\big)
\langle 0|e^{ip_1\cdot \vp_1 }...e^{ip_1\cdot \vp_{M-1}}
e^{ip_M\cdot \vp_M } |0\rangle
+{\rm permutations} \ \cr
&=\sum_{t_1>...>t_{M-1}>0}\ \sum_{x_1,...,x_{M-1}=1}^C
\big( \prod_{i=1}^{M-1}\omega^{-2b_it_i}\big)
\prod_{i<j} e^{-p_i\cdot p_j \langle \vp_i \vp_j \rangle }
+{\rm permutations} \ \ , \cr}
\eqn\amtach
$$
where the chiral correlators are given in eq. \correl\  and the zero
mode as usual yields the energy-momentum conservation delta function.
The fact that only one vertex operator position can be fixed
means, in particular,  that the three-point
function will be non-trivial, i.e. depending on the Mandelstam
invariant $p_1\cdot p_2$, unlike continuum string theory
where the three-point function
on the sphere is a constant (proportional to the string coupling
constant).

Another question concerns whether we should attribute any
physical meaning to the number $C$. In an interacting picture
where strings split and join, a possible requisite is
to demand the interaction to conserve the number of cells, in other
words, that cells are not created or destroyed after each
interaction. This means that when two strings with $C_1$ and $C_2$ cells
join the resulting string would have $C=C_1+C_2$.
In this scenario $C$ would not be a constant of the theory,
but a  large number that may vary,
just as the number of atoms of a macroscopic body.
Clearly, this matter is ignored in the scattering amplitudes
\amplit \ and  \amplito \ where the number of cells has been left
unaltered after consecutive application of vertex operators.
However, even in this scenario, the leading correction
to the continuum string amplitudes can still be obtained from
\amplit\ and \amplito .
Indeed,
we have just argued that the limit $C\to\infty $ exists
and leads to the usual continuum scattering amplitudes. Therefore,
in scattering processes involving strings with  large values of $C$,
any correction to eqs.  \amplit \ and  \amplito \   due to cell number
variation will have to be of subleading order
in a perturbative expansion.

Explicit calculations including the leading corrections to the
scattering amplitudes of the continuum string theory
are under current investigation.
For the moment let us elucidate
the nature of the corrections to the propagator.
The two-point function $\langle \vp \vp\rangle $ is of the form
$$
\sum _{n=1}^N {1\over n}\ x^n
$$
A convenient expression for this sum can be obtained by
writing

$$
\sum _{n=1}^N {1\over n} \ x^n=-\log (1-x) - {x^{\bar N}\over \bar N}
\sum_{n=0}^\infty {1\over 1+{n\over \bar N}} \ x^n\ \ ,
\ \ \ \ \ \bar N\equiv N+1\ .
\eqn\hyperg
$$
By re-arranging the terms in eq. \hyperg\ one obtains
$$
\sum _{n=1}^N {1\over n} \ x^n=-\log (1-x) - {x^{\bar N}\over \bar N}
\sum_{n=0}^\infty \big( {1\over \bar N}\big)^n {d^n\over du^n}
{1\over 1-e^{-u}}\ ,\ \ \ u=-\log x
\eqn\unoene
$$
Thus we see that the standard logarithmic propagator
is corrected by a (non-perturbative) term of order $O(e^{\bar N\log x})$
which multiplies an expansion in powers of ${1\over \bar N}={1\over N+1}$.

It is easy
to prove that the scattering amplitude \amplito\ can also be written as
$$
A^M_{\rm open}={\rm const.}
g^{M-2} \langle 0|W_0^1 \Delta_o W_0^2 \Delta_o ...
W^{M-1}_0 \Delta_o W^M_0|0\rangle +{\rm cyclic \ perm.} \ ,
\eqn\unita
$$
where
$$
\Delta_o=\sum_{t=0}^\infty e^{i(L_0-a)t}\ ,
\eqn\propago
$$
while for closed strings
$$
A^M_{\rm closed}={\rm const.}
\kappa ^{M-2} \langle 0|U_0^1 \Delta_c U_0^2 \Delta_c ...
U^{M-1}_0 \Delta_c U^M_0|0\rangle +{\rm  perm.} \ \ ,
\eqn\ampliclo
$$
where
$$
\Delta_c=\sum_{t=0}^\infty e^{i(L_0+\bar L_0-2a)t}\ ,\ \ \
U_0^i=\sum _{x=1}^C W_{x,0}^i \ \ .
\eqn\propag
$$
In this form  the factorization property is disclosed,
the residue in different channels being
manifestly the product of the corresponding tree amplitudes.

The decoupling of longitudinal states could be alleged on the basis
of the `cancelled-propagator argument': any longitudinal state
can be expressed as a commutator $[H,W']$, where $W'$ is a
lower-level operator, etc. However, a closer inspection shows
that here the mechanisms must be different.
We leave this issue for future work.

\bigskip

\chapter {Deterministic Cellular Strings }

In sect. 3 we promoted $p^\mu $ to an operator in order to apply
the first-quantization procedure. Here we will pursue the pure
cellular
automaton approach a little bit further. The world-sheet variables
of cellular strings are now considered to be fully governed by the
automaton laws of sect. 2; $p^\mu $ is a constant of the two-dimensional
theory that determines how
left and right moving components of the scalar operator transform
under $x\to x+C$.
As discussed at the begining of sect. 3, to make this compatible
with target-space quantum mechanics, one should generalize the techniques
applied on the world-sheet to target space-time;
the physical
states shall be a non-trivial linear combination
of all states of the primitive basis for the Hilbert space
of second quantization, in such a way that their target-space evolution will
look non-deterministic, even though it is deterministic for the
fundamental variables.
\bigskip\bigskip

\noindent 9.1. {\it Deterministic Motion of Free Strings in Target Space}

The existence of a light-cone gauge revealed that the target-space
time $\vp^+ $ has a sort of discrete structure. One can arrive at
a similar conclusion
in the covariant formalism from the ambiguity $\drama $ in the
mass \masasa .
What is more remarkable is
the evolution of  {\it free} cellular strings
 (i.e. $g=0$)  in
$D$-dimensional space-time is deterministic and can be described
by automaton rules.

To be more explicit, let us consider a closed
string and return to the basis in which
$\vp ^\mu $ are diagonal, as described in sect. 2 (see eq. \base )
$$
\{ |v^{0\mu } \rangle \otimes|v_1^{L\mu },...,v^{L\mu }_C\rangle
\otimes |v_1^{R\mu },...,v^{R\mu }_C\rangle \} \ \ .
\eqn\basedos
$$
In this basis we have
$$
\vp^{R\mu }_{x,t}=v^{R\mu }_{x,t}\ ,\ \ \ \ \vp^{L\mu }_{x,t}=v^{L\mu }_{x,t}\
{}.
\eqn\autovalor
$$
It is convenient to work in the light-cone gauge.
Under a tick of the automaton clock, the target time
$\varphi^+=\phi^+ +{\pi l^2\over C} p^+t\ $ changes by
$$
\varphi^+_{t+1}=\varphi^+_t + {\pi l^2\over C} p^+\ ,
\eqn\cambio
$$
and we have the deterministic laws (see eq. \auto )
$$
\varphi_x^{Ri}(\varphi ^+ + {\pi l^2\over C} p^+)
=\varphi _{x-1}^{Ri}(\varphi^+ )\ \ ,
\eqn\autori
$$
$$
\varphi_x^{Li}(\varphi^+ + {\pi l^2\over C} p^+)
=\varphi _{x+1}^{Li}(\varphi^+ )\ .
\eqn\autole
$$

The ``vacuum" $|v=0\rangle $ can be thought of as containing
``hidden variables", being an intrincated superposition
of all possible states (cf. eq. \vcero ).

If $l$ is of the Planck-lenght order, $l\sim l_{\rm P}$,
and $p^+<< 1/l_{\rm P} $, then
$$
\Delta \varphi^+={\pi l^2\over C}p^+
\sim {p^+l_{\rm P}\over C} l_{\rm P}<<l_{\rm P}  .
$$

The evolution of $M$ free closed strings,
$$
\varphi_{x_a}^{ia}=\phi^{ia}+{\pi l^2\over C_a}p^{i}_{a}t_a+\varphi^{Ria}_
{x_a}
+\varphi^{Lia}_{x_a}\ ,\ \ \ i=1,...,D-2\ ,\ \ a=1,...,M\ ,
\eqn\mstrings
$$
is determined in a similar way. The new configuration after a
light-cone time interval $\Delta \varphi^+$ will be
$$
\varphi^{ia}_{x_a} (\varphi^++\Delta\varphi^+)=
\varphi^{Ria}_{x_a-\Delta t_a}(\varphi ^+)
+\varphi^{Lia}_{x_a+\Delta t_a}(\varphi ^+)
+\phi^{ia}(\varphi^+ )+{\pi l^2\over C_a} p^{i}_{a}\Delta t_a\ ,
\eqn\mstringevol
$$
$$
\Delta t_a=[[{C_a\over \pi l^2 p^{+}_{a} }\Delta\varphi^+ ]]\ ,
$$
where $[[...]]$ denotes integer part.

\bigskip\bigskip\bigskip
\noindent 9.2. {\it Interacting Strings -- A Model}

An important
question is whether there exist deterministic automaton rules
dictating the evolution and interaction
of many strings which reproduce the behaviour of usual continuum strings at
long-distance scales.
In our view, there is no reason to think that this
is impossible; for example, in refs. [\hooft,\hooftt ]
it was shown that extremely simple
automaton rules can lead to chaotic behaviour.
It is conceivable that the cognition of a hypothetical
second-quantized string theory
should uncover such deterministic laws.

To illustrate the evolution of interacting strings, let
us consider a simple example. A possible (cell conserving)
interacting rule is the following.

A string $1,...,C$ with variables evolving
according to eqs. \lawzero, \autom\ or eqs. \derecha -\bautopen  \
 --depending
on whether the string is closed or open--  breaks
if at some time there is a cell $y$ where the following
inequality is satisfied
$$
{\cal E}_{y,y+1}\equiv (v^L_{y,t}-v^L_{y+1,t})^2
+(v^R_{y,t}-v^R_{y+1,t})^2 \geq K\ l^2\ \ ,
\eqn\splitting
$$
where  $K$ is a given positive real number, related to the string
coupling constant. If the string was closed, the result of the breaking
is
an open string with ends at the cells $y$ and $y+1$, i.e. the cell string
$\{ y+1,y+2,...,C,1,...,y\} $ (see fig. 1).

If the string was open, two open strings with
$C_1=y$ and $C_2=C-y$ cells will emerge (fig. 2).
A natural rule prescribing the values of
the center of mass momenta carried by each of the resulting strings
is attainable by summing up the `momenta' associated with the
individual cells.
Let us consider an open string with cells $0,1,...,N+1$.
In analogy with the continuum theory, the momentum carried by a
single cell is
$$
\eqalign{
\delta p_{x,t}^\mu &={1\over \pi l^2}
(\varphi ^\mu_{x,t}-\varphi^\mu_{x,t-1}) \cr
 &={1\over N+1}p^\mu +{1\over \pi l^2}(v^{R\mu}_{x,t}-v^{R\mu}_{x,t-1}+
v^{L\mu}_{x,t}-v^{L\mu}_{x,t-1}) \ ,\ \ \ x=1,...,N \ .\cr}
\eqn\delpe
$$
The momentum that should be assigned to the end cells can be found
by demanding
$$
\sum_{x=0}^{N+1}\delta p_{x,t}^\mu = p^\mu ={\rm constant}
\eqn\costante
$$
According to the evolution rules \autopen\ and the boundary condition
\bautopen\ the quantity
$$
{\cal V}^\mu \equiv v^{L\mu}_{0,t} + v^{R\mu}_{N+1,t}
+\sum_{x=1}^N (v^{R\mu}_{x,t}+v^{L\mu}_{x,t})
\eqn\vvvv
$$
is time-independent, as it can be easily verified. This suggests us
to define the end cell momenta as
$$\eqalign{
\delta p^\mu_{x,t}&=
{1\over 2N+2}p^\mu +{1\over 2\pi l^2}\big( v^{R\mu}_{x,t+1}-v^{R\mu}_{x,t}+
v^{L\mu}_{x,t+1}-v^{L\mu}_{x,t} \big)\cr
&={1\over 2N+2}p^\mu +{1\over \pi l^2}\big(
v^{R\mu}_{x,t+1}-v^{R\mu}_{x,t}\big) \ ,\ \ \ x=0,N+1\ ,
\cr}
\eqn\endmomen
$$
i.e., half the expression corresponding to a non-extremal cell, eq.\delpe .
{}From eq.\endmomen\ it follows
$$
\sum_{x=0}^{N+1}\delta p_{x,t}^\mu = p^\mu +
{1\over \pi l^2}\big( {\cal V}_{t}^\mu - {\cal V}_{t-1}^\mu\big) =p^\mu\ ,
$$
as desired.

An alternative reflecting boundary condition, which differs from eq.
\bautopen\ by a delay in the reflection process of one time step, is
the following:
$$
v^{R\mu}_{x, t+1}=v^{R\mu}_{x-1, t}\ ,\ \ \
v^{L\mu}_{x, t+1}=v^{L\mu}_{x+1, t}\ ,\ \ \ \ x=0,1,...,N+1\ \ ,
$$
$$
v^{R\mu }_{-1,t}=v^{L\mu}_{0,t}\ ,\ \ \ \
v^{L\mu }_{N+2,t}=v^{R\mu}_{N+1,t}\ .
\eqn\abautopen
$$
Then the eq. \delpe\ for $\delta p_{x,t}^\mu$
as well applies to the end cells. The
time-independent quantity is now
$$
{\cal V}^{\mu} \equiv \sum_{x=0}^{N+1} (v^{R\mu}_{x,t}+v^{L\mu}_{x,t})
\eqn\avvvv
$$
Although both boundary rules
give rise to the same theory in the
continuum limit, the boundary condition
\abautopen\ may not be very convenient in the first-quantization
treatment, since it
yields a relative phase factor in the formula connecting
left and right mode operators, $\bar\ga_n ^\mu =\xi \ga^\mu_n $.

Now we are ready to pronounce the deterministic rule giving
the distribution of momenta of the emerging strings.
At the instant of the splitting of the open string
$1,...,C$ into the open strings I=$1,...,y$ and
II=$y+1,...,C$, the total momentum carried by string I is
$$\eqalign{
p_{\rm I}^\mu=\sum_{x=1}^y\delta p^\mu_x &= {y\over C-1} p^\mu +
{1\over \pi l^2}\big( {\cal V}_t^\mu -{\cal V}_{t-1}^\mu\big) \cr
&={y\over C-1} p^\mu+ {1\over \pi l^2}\big( v^{L\mu}_{y,t} -
v^{R\mu}_{y+1,t} \big) \ .\cr}
\eqn\viauno
$$
Similarly
$$
p_{\rm II}^\mu =\sum_{x=y+1}^C \delta p^\mu_x = {C-y\over C-1}
p^\mu +{1\over \pi l^2}\big( v^{R\mu}_{y+1,t} -v^{L\mu}_{y,t} \big) \ .
\eqn\vialaltro
$$
In particular, note that energy-momentum is conserved,
$$
p^\mu=p_{\rm I}^\mu+p_{\rm II}^\mu\ .
$$

Now let us consider joining of strings.
The two end cells $x_1$ and $x_2$ of an open string join
if the corresponding
${\cal E}_{x_1,x_2}$ becomes lower than $K \ l^2$, i.e.
$$
{\cal E}_{x_1,x_2}\equiv (v^L_{x_1,t}-v^L_{x_2,t})^2
+(v^R_{x_1,t}-v^R_{x_2,t})^2 < K\ l^2 \ .
\eqn\joining
$$
The result is a closed string with the same value of momentum $p^\mu $.

In stating the rule for joining of end cells belonging
to distinct strings we have to take into account the
role of the center of mass coordinate $v^{0a}_{t_a}$, where $a$ labels
the different strings. Let us normalize the variables
$v^{L\mu a}_x, v^{R\mu a}_x$
by adding them a global constant such that ${\cal V}^{\mu a}=0$.
Then we declare
that the end cells $x_1,x_2$ of string I and string II join if
$$
{\cal E}_{x_1,x_2}\equiv (v^{0{\rm I}}_{t_1}-v^{0{\rm II}}_{t_2})^2+
(v^{L{\rm I}}_{x_1,t_1}-v^{L{\rm II}}_{x_2,t_2})^2
+(v^{R{\rm I}}_{x_1,t_1}-v^{R{\rm II}}_{x_2,t_2})^2 < K\ l^2 \ .
\eqn\gjoining
$$
where
$$t_{1,2}=[[{C_{\rm I,II}-1\over \pi l^2 p^{+}_{\rm I,II} }
\big( \varphi^+ -\phi^{+\rm I,II}\big) ]]\ ,\ \ \
v^{0i{\rm I,II}}_{t_{1,2}}=\phi^{i{\rm I,II}} + {\pi l^2\over C_{\rm I,II}-1}
p^i_{\rm I,II}t_{1,2}\ ,
$$
and $C_{\rm I, II}$ are, respectively, the number of cells
of strings I and II (cf. eqs. \vpopen ).
The result will be an open string with
$C_{\rm I}+C_{\rm II}$ cells.

After each splitting or joining of the strings, the evolution
of each of the resulting strings
is afresh dictated by eqs.
\lawzero, \autom\ or eqs. \derecha -\bautopen  \
until another splitting or joining configuration is encountered.

Since closed strings can turn into open strings and viceversa
these rules prescribe a theory which necessarily includes both
open and closed (oriented) strings.

Finally, the splitting
of a closed string into two closed strings
(and similarly for the joining process) is performed
when a `double' configuration is met, as depicted in fig. 3.
There are two (non-consecutive) cells $x$ and $y$
where eq. \splitting\ is satisfied,
${\cal E}_{x,x+1}\geq K\ l^2$, ${\cal E}_{y,y+1}\geq K\ l^2$,
and at the same time
${\cal E}_{y,x+1}<K\ l^2$, ${\cal E}_{x,y+1}< K\ l^2$.
Another situation is that the splitting of a closed string
into two (or more) closed strings occurs in a finite number
of automatom time steps, i.e. by first breaking into two
(or more) open strings whose ends join in succesive steps (see fig. 4).
In the continuum limit the process will appear to happen
instantaneously; it
should be indistinguishable from an elementary process.

The parameter determining the interaction
${\cal E}_{x_1,x_2}$ has a simple geometric interpretation
in the space of the automata variables
as the square distance
between the points $\{ v^{0{\rm I}}, v^{Li{\rm I}}_{x_1} , v^{Ri{\rm I}}
_{x_1} \} $
and $\{ v^{0{\rm II}}, v^{Li{\rm II}}_{x_2}, v^{Ri{\rm II}}_{x_2} \} $.
Intuitively, the string breaks because when ${\cal E}_{x,x+1}\geq K\ l^2$
`it costs too much energy' for the string to keep the cells
$x$ and $x+1$ together (in fact, in the continuum limit ${\cal E}_{x,x+dx}$
is the energy associated with an infinitesimal segment $dx$).

Let us consider the evolution of a single closed string.
If the initial configuration is such that ${\cal E}_{y,y+1}$
is $<< K\ l^2$ for all $y$, then the values of ${\cal E}_{y,y+1}$
will slightly oscillate in time but will never reach
the critical value $K\ l^2$, so the string will propagate
without ever breaking.
If, instead, the initial configuration has some ${\cal E}_{y,y+1}$
near the critical value, then splitting may occur during the
automaton evolution.
Similarly, given $M$ (open or closed) strings, there are initial
configurations such that
they will propagate forever without any interaction, but in a generic case
splitting and joining will occur. The interaction rate is adjustable
by varying $K$.

We have tested these rules by computer simulations
and obtained reasonable behaviour.
A thorough study is necessary to decide
whether this interacting cellular string
theory approximates the dynamics of continuum strings
in the long-distance regime.

\bigskip

\chapter {Conclusions}

The fundamental physical principles of string theory are presently
unknown. Considerable progress has been achieved in various areas,
but crucial conceptual problems, such as e.g. the off-shell extension,
still remain in an opaque and puzzling status.
The presence of a Hagedorn phase transition in
string theories was interpreted as a limitation in the range of their
applicability. Several authors advocated the necessity
of a more fundamental theory. In view of the fact that the Hagedorn
transition is triggered by a condensation of punctures in the Riemann
surface, it was speculated that such a fundamental theory
should be formulated on discrete world sheets [\aw ].
However, there are numerous theoretical
constraints that a theory of this sort should satisfy.
Here we have investigated a `minimal' theory which seems to
meet all the required properties.
In addition, it might hopefully
lead to a widely yearned-for result,
namely the restoration of determinism in the evolution
at Planckian scales.

We have seen, nonetheless, that a discrete world-sheet is intimately
linked to a certain discrete target space-time structure.
More precisely, we have seen that a discrete world-sheet time
implies (in the light-cone gauge) that also the target light-cone time has
a discrete structure,
leading inevitably to relations of the form \drama\ .
The problem involved in this relation is the infinite
degeneracy of the mass spectrum in virtue of an infinite tower
of states for each mass.
It is very unclear whether this property represents
a problem for the construction of realistic models.
Presumably, below the Hagedorn phase transition discrete strings
can be effectively replaced by continuum strings and hence
these physical states acquire  large masses.
In any case, the model presented here is relatively simple.
 {\it A priori}, other more complex possibilities
like, for example, random fractional time steps, or {\it non-rigid} time steps,
in which the lapse of the
time step depends on the values of the local variables,
cannot be excluded.

Though unavowed in sect. 2,
determinism in the evolution of world-sheet variables is still
present in the $N\to \infty $ continuum limit, but it no longer
admits an interpretation in terms of a cellular automaton.
Deterministic systems are found wherever there is a basis in which
the wave-function does not spread [\hooft ].

We have deliberately separated the issue of
discreteness of the world sheet from
the issue of determinism in the evolution, since, in principle,
 the former does not
necessitate the latter, though it may be thought of as
a prior instance.
World-sheet discreteness leads to
an indefinition of the structure constants of the algebras we studied,
which translates into
ambiguities in physical quantities, such as the square mass
of the physical states or the norm of Fock space states.
As mentioned in sect. 3, this problem is absent in the fundamental
system in terms of automaton variables. The problem arises upon
introducing the Hilbert space extension of the system, which
is needed to make contact with the usual continuum physics.
In sect. 8 we discussed a prescription to remove such ambiguities
in the continum limit, but the underlying physical mechanisms
are unclear.

We have seen that free cellular strings evolve in space-time
governed by deterministic rules and speculated that there should
as well exist conceivable deterministic rules dictating the evolution of
interacting strings. A concrete model was introduced for
a string theory where the full evolution, including splitting
and joining of an arbitrary number of closed and open cellular strings,
is deterministic.
While the interaction
rules we propose are quite simple, the resulting string dynamics
is highly complicated insofar as
its description requires numerical calculation or
statistical methods.

The reconciliation with target-space
quantum mechanics {\it \` a la} 't Hooft
requires a second-quantization framework.  One has to define
a proper Hilbert space
of all possible string states admitting arbitrary occupation numbers,
then relate the physical states to the primitive basis, etc.
In the context of the interacting model of sect. 9.2, the
string amplitudes may be  regarded just a first-quantization
method to obtain quantum mechanical probabilities,
 but the inexorable future of the system will have been
pre-established at the very moment the initial configuration
was given.

There are a number of technical and conceptual points which need
detailed investigation.
In particular, the properties of the scattering
amplitudes  are yet to be understood. Of course, it would be
inadequate to devote much attention to scattering amplitudes
in a system which aspires to be deterministic, but it is necessary
to establish a more precise connection with the continuum theory.
For the same reason, it may be worth to develop
a higher-genus formulation.

Finally, it would also be interesting to consider
possible applications of the ${\bf Z}_C$-string theory to QCD.

\bigskip\bigskip
 \noindent $\underline {\rm Acknowledgements}$: The author is grateful
to L. Susskind for a stimulating conversation and L. Rozansky for
a useful remark.

 \refout
\end